\documentclass[11pt,final,onecolumn]{IEEEtran}
\usepackage{hyperref}
\usepackage{graphicx}
\usepackage[cmex10]{amsmath}
\usepackage{amsthm}
\usepackage{amssymb}
\newcounter{MYtempeqncnt}
\begin{document}

\title{The Stationary Phase Approximation, Time-Frequency Decomposition and Auditory Processing}

\author{Bernard Mulgrew, \textit{Fellow IEEE}\\
    Institute for Digital Communications\\
    School of Engineering\\
    The University of Edinburgh\\
    Edinburgh, EH9 3JL, UK\\
    Email: B.Mulgrew@ed.ac.uk\\\thanks{This work was supported by the Royal Academy of Engineering and SELEX Galileo. Submitted to IEEE Trans Signal Processing 14th Aug 2012}}

\maketitle
\IEEEpeerreviewmaketitle

\begin{abstract}
The principle of stationary phase (PSP) is re-examined in the context of linear time-frequency (TF) decomposition using Gaussian, gammatone and gammachirp filters
at uniform, logarithmic and cochlear spacings in frequency.
This necessitates consideration of the use the PSP on non-asymptotic integrals and leads to the introduction of a test for phase rate dominance. Regions of the TF plane that pass the test and don't contain stationary phase points contribute little or nothing to the final output. Analysis values that lie in these regions can thus be set to zero, i.e. sparsity. In regions of the TF plane that fail the test or are in the vicinity of stationary phase points, synthesis is performed in the usual way. A new interpretation of the location parameters associated with the synthesis filters leads to: (i) a new method for locating stationary phase points in the TF plane; (ii) a test for phase rate dominance in that plane.
Together this is a TF stationary phase approximation (TFSFA) for both analysis and synthesis.
The stationary phase regions of several elementary signals are identified theoretically and examples of reconstruction given.
An analysis of the TF phase rate characteristics  for the case of two simultaneous tones predicts and quantifies a form of simultaneous masking similar to that which characterizes the auditory system.
\end{abstract}

\begin{IEEEkeywords}
method of re-assignment; cochlear filters; gammatone; gammachirp;  simultaneous masking;
\end{IEEEkeywords}


\section{Introduction}
\label{sec:intro}
The principle (or method) of stationary phase (PSP) \cite{Wong1989book} is a result from asymptotics that can provide closed-form approximations, in the limit as $\lambda \rightarrow \infty$,  to often intractable oscillatory integrals of the form
\begin{eqnarray} \label{eqn:complex_integral}
  I &=& \int_{- \infty}^{\infty} \! a(t) \mathrm{e}^ {\jmath \lambda b(t)} \mathrm{d}t
\end{eqnarray}
where $a, b, t, \lambda \in \mathbb{R}$.
There are two steps involved: (i) recognition that in the limit the integral will be almost zero everywhere in the interval $\{ t: - \infty \leq t \leq \infty \} $ except near values of $t$ where the derivative $\dot{b}(t)$ is zero, the stationary phase points; (ii) the integral in the vicinity of these stationary phase points can be expressed in terms of the second derivative of the phase i.e. $\ddot{b}(t)$. Perhaps the most successful application of the PSP in signal processing has been in the context of synthetic aperture radar (SAR), where it is the starting point in the development of many of the Fourier-based imaging algorithms, c.f. \cite{SoumekhBook1999}.
Application of the PSP in not without its pitfalls.
It is tempting to use the PSP in the non-asymptotic cases where $\lambda = 1$ to find closed form approximations to integrals such as Fourier transforms.  The argument for this requires that the phase $b(t)$ is changing much more rapidly than the amplitude $a(t)$.
However, as pointed out in \cite{Chassande1998TFTSA}, a degree of care must be exercised, particularly with step (ii).

The primary interest here is its application to linear time-frequency (TF) decomposition \cite{MallatBook2nd}.
The motivation is the recent resurgence of interest in  analogue filter banks both as part of a synthetic cochlea and as a means to provide power efficient implementations of analysis filter banks \cite{Mandal2009SSC}.
The desire with both is to extract salient features from the TF decomposition using the limited functionality associated with analogue circuitry.
This does not deny the considerable work that have been done on computational modelling of the auditory system, typified by papers such as \cite{jepsen:422} and the references therein. However the main purpose of such work is to model and predict the response of the auditory system to stimulus rather than to
expose the signal processing principles that might be at work.

The PSP is a natural place to start because of the prevalence of oscillatory terms in TF decompositions and the hope that the stationary phase points may provide a means for identifying salient features as well as a focus for sparse decompositions without the need for the usual iterative re-synthesize  steps \cite{Chen1998SIAM}.
The PSP has been applied to linear TF decomposition for both  analysis, \cite{Delprat1992IT} \& \cite{Guillemain}, and  synthesis \cite{Kodera1978TSP}, the latter leading to the method of reassignment.
The approach here is to revisit \cite{Kodera1978TSP} and to fundamentally re-interpret it
to provide a PSP-based approximation
to the TF synthesis integral.
There is no attempt to either reassign \cite{Kodera1978TSP} or relocate \cite{Daubechies2010ACHA} components in the TF plane because of the limited functionality mentioned above.
Another alternative would be to follow an amplitude-based approach such as ridgelets \cite{Carmona1999TSP}. However this leads to algorithms that are far from feasible with analogue circuitry and further, ridges in the TF plane may not be appropriate when dealing with auditory filters such as the gammatone \cite{Lyon2010ISCAS} and gammachirp \cite{Irino1997JASA} which have asymmetrical impulse responses and, in the case of the latter, an asymmetrical frequency response.

Concentration on the synthesis rather than the analysis integral is advocated because:
(i) most methods for sparse atomic compensation, c.f. \cite{Chen1998SIAM}, are based on re-synthesis of the original waveform;
(ii) one of the main functions of the auditory system is to code the incident waveforms
and coding requires at least some consideration of the potential for reconstruction (even when reconstruction is not a requirement);
(iii) smooth variations of the magnitude and phase of the integrand are more readily satisfied for the synthesis integral than the analysis integral (because the signal has already been filtered in the analysis process).

In Section \ref{sec:prelim} a preliminary description of the filter banks to be considered is given. A framework is presented that accommodates Gaussian, gammatone and gammachirp filters at uniform, logarithmic and cochlear spacings in frequency. More details of the relevant properties are provided in Appendix \ref{app:proto}.
Section \ref{sec:PSFNAI} is an examination of the application of the PSP to non-asymptotic integrals and proposes the use of step (i), above, without step (ii). The concept of phase-rate dominance is introduced to partition the interval into sets where the PSP is and is not applied.
The synthesis double-integral is addressed in section \ref{Sec:sparsity} and the PSP is applied to it in a new way to facilitate the detection of stationary phase points in the TF plane. Extending the single-integral concepts to double integrals (Appendix \ref{app:PSFNAI2D}) leads to a test for phase-rate dominance in the TF plane.
The stationary phase regions of several elementary signals are identified in section \ref{sec:elementary} and examples of reconstruction provided.
Section \ref{sec:simultaneous} examines the phase rate characteristics in the TF plane for the case of two simultaneous tones . It is shown that if these characteristics are used to test for the presence of a tone at a particular frequency, a form of simultaneous masking, similar to that which characterizes the auditory system \cite{MooreBook2003}, is observed.
All of the above is addressed from the perspective of deterministic signals. Consideration of random processes is left for future work. The effects of noise on the related  method of reassignment are covered in  \cite{Gardner2006PNAS} and\cite{Miroslav201177}.
Finally, in section VII, conclusions are drawn. 
\section{Preliminaries}\label{sec:prelim}
Consider a time-frequency analysis $X_{\omega} (t)$ of a signal of
interest $x(t)$ of the form:
\begin{eqnarray}\label{eqn:analysis}
    X_{\omega} (t) = x(t) * h_{\omega} (t), \,\,\, \{  \omega : \omega_{min} < \omega < \omega _{max} \}
\end{eqnarray}
where $*$ denotes convolution and the impulse response of a single
filter in the analysis filter bank is given by:
\begin{eqnarray}\label{eqn:filter}
  h_{\omega} (t) &=& \beta h(\beta t) \mathrm{e}^{\jmath \omega t}
\end{eqnarray}
Each filter is formed using a prototype filter $h(t)$ and a nominal bandwidth $\beta$.
The frequency response of the analysis filter is related to the frequency response of the prototype in a straight forward way, i.e. $H_{\omega} (\Omega) = \textsf{F} [h_{\omega} (t)]  = H    \left ( \frac{ \Omega - \omega}{\beta} \right )$.
where $H(\Omega) = \textsf{F} [h (t)] = \int_{-\infty}^{\infty} \! h(t) \mathrm{e}^{- \jmath \Omega t} \, \mathrm{d} t$.
Prototype filters that will be considered here are
a gammatone pulse, a gammachirp pulse and a Gaussian pulse. While the Gaussian pulse is a common \cite{MallatBook2nd} and analytically convenient choice for time-frequency analysis, the gammatone \cite{Lyon2010ISCAS} and gammachirp pulses \cite{Irino1997JASA} more closely model the cochlea in the ear.
Details of these
prototype filters and their properties are summarized in
Appendix \ref{app:proto}.
The prototype filters are
normalised such that $H(0) = 1 $. This
property and multiplication by the nominal bandwidth $\beta$ in (\ref{eqn:filter})
normalizes the maximum gain of each filter to unity at $\omega$ rad/s. A re-synthesis, $\hat{x}(t)$,  of the signal of
interest is performed using filters matched to $h_{\omega} (t)$. Thus for real signals of interest
\begin{eqnarray}\label{eqn:synthesis_integral}
  \hat{x}(t) &=& \frac{1}{C} \Re \left \{ \int_{0}^{1} \! \! \int_{-\infty}^{\infty} \! Z(\tau , \mu) \, \mathrm{d}\tau   \mathrm{d}\mu \right \}
\end{eqnarray}
with integrand $Z(\tau , \mu)= X_{\omega}(\tau) h_{\omega}^{*} (\tau-t)$
and where $C$ is a constant and $\Re \{.\}$ and $\Im\{.\}$ denote `real part of' and `imaginary part of' respectively.
As a convenience, in order to deal with a number of possible filter bank spacings, define a filter bank variable $\mu$ that lies in the range $[0, \, 1]$, A value $\mu = 0$ indicates the lower edge of the filter bank and $\mu=1$ indicates the upper edge. The  frequency $\omega$ at which the filters gain is a maximum is a function of $\mu$ and the bandwidth of the filter $\beta$ is proportional to the derivative $\frac{\mathrm{d} \omega}{\mathrm{d} \mu}$, i.e. $\beta \propto \frac{\mathrm{d} \omega}{\mathrm{d} \mu}$.
Hence the total number of filters required to just cover the band of interest is given approximately by $\frac{\mathrm{d} \omega}{\mathrm{d} \mu} \frac{1} {\beta}$.
In the following $\omega_{min}$ is nominally the lowest frequency covered by the filter bank and $\omega_{max}$ is the maximum.
For a uniformly spaced filter bank $\omega = \{\omega_{max} - \omega_{min}  \} \mu + \omega_{min}$
and hence $u = \{\omega - \omega_{min} \}/\{\omega_{max} - \omega_{min} \}$,
$\frac{\mathrm{d}\omega}{\mathrm{d}\mu} =  \omega_{max} - \omega_{min}$,
the nominal bandwidth $\beta$ is a constant and $\frac{\mathrm{d}\beta}{\mathrm{d}\omega} = 0$.
For logarithmically spaced filter banks, similar to wavelets \cite{MallatBook2nd}, $\omega = \omega_{min} \mathrm{e} ^{b \mu}$
where $b = \ln (\frac{\omega_{max}}{\omega_{min}})$ and hence $\mu = \frac{1}{b}  \ln \left ( \frac{\omega}{\omega_{min}} \right )$, $\frac{\mathrm{d}\omega}{\mathrm{d}\mu} = b \omega$, the nominal bandwidth $\beta$ is proportional to $\omega$ and $\frac{\mathrm{d}\beta}{\mathrm{d}\omega}$ is a constant.
For a cochlear spaced filter banks, based on the approximation of \cite{Moore1987209} for low sound pressure levels,
$\omega  = \{  \omega_{min} +a \} \mathrm{e}^{b \mu} - a$,
where $a = \frac{2 \pi \times 10^3}{4.37}$,  $b = \ln \left ( \frac{ \omega_{max} + a}{ \omega_{min} +a } \right )$ and hence $\mu = \frac{1}{b}   \ln \left ( \frac{ \omega + a}{ \omega_{min} +a } \right )$, $\frac{\mathrm{d}\omega}{\mathrm{d}\mu}  =  b \{  \omega + a \}$ and $ \frac {\mathrm{d} \beta}{\mathrm{d}\omega} $ is a constant.
For all the above filter banks, the equivalent rectangular bandwidth (ERB) \cite{MooreBook2003} and the 3 dB bandwidth are related to the nominal bandwidth $\beta$ in a straightforward way.

\section{The PSP and non-asymptotic integrals}
\label{sec:PSFNAI}
Consider the integral (\ref{eqn:complex_integral}) evaluated over an interval $[t - \Delta/2, t + \Delta /2 ]$. 
Assuming that the function $f(t) = a(t) \mathrm {e}^{\jmath b(t)}$ is well approximated by its Taylor series over this interval, gives
\begin{eqnarray}
\nonumber
  I_{\Delta} (t)&\triangleq& \int _{t - \Delta/2}^{t + \Delta /2 } \, f(t^{\prime}) \, \mathrm{d}t^{\prime} \\
\nonumber  
  &\approx& f(t) \int_{t-\Delta /2}^{t+\Delta /2}  \! { \textstyle \left \{ 1 + \frac{\dot{f} (t)}{f(t)} \{ t^{\prime} - t\} + \frac{\ddot{f}(t)}{f(t)} \frac{\{ t^{\prime} - t\} ^{2}}{2} \right \} } \, \mathrm{d}t^{\prime}
\end{eqnarray}
The derivatives can expressed as 
\begin{eqnarray}
\nonumber
   \frac{\dot{f}(t)}{f(t) }&=&  \left \{ \frac{\dot{a}(t)}{a(t)} + \jmath \lambda \dot {b} (t) \right \}
\end{eqnarray}
and:
\begin{eqnarray}
\nonumber
  \frac{ \ddot{f}(t)}{f(t)} &=&  \frac{\ddot{a}(t)}{a(t)} - \left \{ \frac{\dot{a} (t)}{a(t)} \right\}^ {2}  + \jmath \lambda \ddot{b}(t)  +  \left \{ \frac{\dot{a}(t)}{a(t)} + \jmath \lambda \dot {b} (t) \right \}^{2}
\end{eqnarray}
In the asymptotic case, at a suitably large value of $\lambda$,  $\frac{\dot{f}(t)}{f(t)} \approx  \jmath \lambda \dot {b} (t) $ and $\frac{\ddot{f}(t) }{f(t)} \approx -  \lambda^{2} \dot{b}^{2} (t) + \jmath \lambda \ddot{b} (t)$
and hence the integral is dominated by the phase derivatives $\dot{b} (t)$ and $\ddot{b} (t)$. Thus, at a stationary point, where $\dot{b} (t)=0$, the integral can be evaluated in terms of $\ddot{b} (t)$ without reference to derivatives of $a(t)$.
For the non-asymptotic case where $\lambda = 1$, these the approximations are also dependent on the relationships between the derivatives of the magnitude and phase of the integrand. Thus $\frac{\dot{f}(t)}{f(t)} \approx \jmath  \dot {b} (t)$
provided, as in \cite{Delprat1992IT}, that
\begin{eqnarray}\label{eqn:PSF_inequality1}
  \mid \dot {b} (t) \mid  & \gg & \left | \frac{\dot{a}(t)}{a(t)} \right  |
\end{eqnarray}
and $\frac{\ddot{f}(t)}{f(t)}
   \approx   -   \dot{b}^{2} (t) + \jmath  \ddot{b} (t)$
provided \textit{also} that
\begin{eqnarray} \label{eqn:PSF_inequality2}
    \mid \ddot{b} (t) \mid & \gg & \left | \frac{\ddot{a}(t)}{a(t)} - \left \{ \frac{\dot{a} (t)}{a(t)} \right\}^ {2}  \right |
    = \left | \frac{\mathrm{d}}{\mathrm{d}t} \left ( \frac{\dot{a} (t)}{a(t)} \right ) \right |
    \end{eqnarray}
If the amplitude and phase derivatives are available then at every value of $t$ it is possible to test for what might be called \textit{first order or second order phase-rate dominance} using (\ref{eqn:PSF_inequality1}) or, (\ref{eqn:PSF_inequality1}) and (\ref{eqn:PSF_inequality2}), respectively.
The approach adopted here, where a closed form approximation to the integral is not required, is to  circumvent the difficulties associated with step (ii) of  the PSP by simply not using that step. Rather (\ref{eqn:PSF_inequality1}) is used to identify a set $S_{0}$ that is the union of intervals of $t$ where step (i) is valid and the compliment to that set $\bar{S}_{0}$ where it is not. If the stationary phase points $\{ t_{i}\}_{i}$ where $\dot{b}(t_{i})=0$ and intervals such as $S_{i} \in \{t : t_{i} - \delta_{1} < t < t_{i} + \delta_{2}, \delta_{1} \geq 0, \delta_{2} \geq 0  \}$ that contain them can be identified, the integral over the whole real line  (\ref{eqn:complex_integral}) can be replaced by a similar integral over the union of intervals $S \in \{\bigcup _{i} S_{i} \} \cup \bar{S}_{0}$.
It is also worth noting that
for (\ref{eqn:PSF_inequality1}) to be satisfied at or near a stationary phase point, would also require that $| \frac{\dot {a} (t) |} {a(t)}| \rightarrow 0  $ as $t \rightarrow t_{0}$. Thus the normalized amplitude rate must go to zero more rapidly than the phase rate, i.e. (\ref{eqn:PSF_inequality2}) must apply.
Thus there are liable to be be intervals where $S_{i} \in \bar{S}_{0}$ and for these intervals there is no need to identify $\delta_{1}$ and $\delta_{2}$.

\section{A Time-Frequency Stationary Phase Approximation}
\label{Sec:sparsity}

The PSP can be extended to double integrals such as (\ref{eqn:synthesis_integral}) with the result defined in terms of the gradient and Hessian of the phase of the integrand, c.f. \cite{Wong1989book}, pg. 478.
However, as in Section \ref{sec:PSFNAI} the full form the PSP is not used here. Specifically, the gradient is used in step (i) to identify the stationary phase points but the Hessian of step (ii) is not used to form an approximation to the integral in the vicinity of these points. A test similar to (\ref{eqn:PSF_inequality1}) is developed for double integrals in Appendix \ref{app:PSFNAI2D}
to identify regions of phase-rate dominance in the TF plane where the PSP is applied by numerical integration in the vicinity of the stationary phase points.
As in Section \ref{sec:PSFNAI}, it is convenient to define  normalized time- and frequency- derivatives of the integrand $Z(\tau, \mu)$, i.e. $Z_{\tau} \triangleq \frac{ \frac{\partial }{\partial \tau}Z (\tau, \mu)}{Z(\tau, \mu)}$ and $Z_{\mu} \triangleq \frac{ \frac{\partial }{\partial \omega}Z (\tau, \mu) \frac{d \omega}{d \mu}}{Z(\tau, \mu)}$ respectively, where
\begin{eqnarray}\label{eqn:Z_tau}
\nonumber
Z_{\tau}
 &=& \frac{\frac{\partial }{\partial \tau}X_{\omega}(\tau)}{X_{\omega}(\tau)} + \frac{\frac{\partial }{\partial \tau}h_{\omega}^{*} (\tau-t)}{h_{\omega}^{*} (\tau-t)}\\
   &=& \frac{\frac{\partial }{\partial \tau}X_{\omega}(\tau)}{X_{\omega}(\tau)} + \beta \frac{{\dot{h}^{*}} (\beta \{ \tau-t\})}{h^{*} (\beta \{ \tau-t\})} - \jmath \omega
\end{eqnarray}
and
\begin{eqnarray}\label{eqn:Z_mu}
Z_{\mu}
&=& \left \{ \frac{\frac{\partial }{\partial \omega}X_{\omega}(\tau)}{X_{\omega}(\tau)} + \frac{\frac{\partial }{\partial \omega}h_{\omega}^{*} (\tau-t)}{h_{\omega}^{*} (\tau-t)} \right \} \frac{\mathrm{d}\omega}{\mathrm{d}\mu}
\end{eqnarray}
where
\begin{eqnarray}\label{eqn:Z_mu_extra}
\nonumber
\lefteqn { \frac{\frac{\partial }{\partial \omega}h_{\omega}^{*} (\tau-t)}{h_{\omega}^{*} (\tau-t)}   } \\
\nonumber
& & = \,\, \frac{\mathrm{d}\beta}{\mathrm{d}\omega} \left \{ {  \frac{1}{\beta} +  \frac{{\dot{h}^{*}} (\beta \{ \tau-t\})}{h^{*} (\beta \{ \tau-t\})} \{ \tau -t \}  } \right \}  - \jmath \{\tau -t \}\\
& &
 \end{eqnarray}
Both $Z_{\tau}$ and $Z_{\mu}$ are additions of a signal dependent term, e.g. $\frac{\frac{\partial }{\partial \tau}X_{\omega}(\tau)}{X_{\omega}(\tau)}$, and a filter dependent term, e.g. $\frac{\frac{\partial }{\partial \tau}h_{\omega}^{*} (\tau-t)}{h_{\omega}^{*} (\tau-t)}$. The former is  a function of the pair $(\omega , \tau)$ whereas the latter is a function of the pair $(\omega, \tau - t)$.  Note that $(\omega, \tau - t)$ are themselves parameters of the filter, specifically, the frequency where the filter has maximum gain $\omega$ and the delay $\tau -t$ between the input and output of the filter.
In contrast to the method of
re-assignment derived in \cite{Kodera1978TSP}, the delay term $\{\tau - t \}$ is interpreted as
the group delay
\begin{eqnarray} \label{eqn:group_delay}
 \{\tau - t \}  =   g( \omega ) & \triangleq &  - {\textstyle \frac{\mathrm{d} }{\mathrm{d} \Omega}} \angle H_{\omega} (\Omega) {\big |} _{\Omega = \omega}
\end{eqnarray}
of the filter (\ref{eqn:filter}) at $\omega$, where the notation $\angle H$ indicates the argument of the complex variable $H$.
Simple expressions for the group delay are given in the Appendix \ref{app:proto} in
terms of the group delays of the prototype filters at zero
frequency. The justification for the use of (\ref{eqn:group_delay}) in
(\ref{eqn:Z_tau}),
(\ref{eqn:Z_mu})
and
(\ref{eqn:Z_mu_extra})
proceeds as follows:
$\{ \tau - t \}$ is the delay between the input signal $x(t)$ and the output of the analysis filter $X_{\omega}(\tau)$; since each filter is tuned to have a maximum gain a particular frequency $\omega$, the delay through the filter is also tuned to the rate of change of the phase response at the frequency where the gain is maximum. A particular filter is thus jointly labeled with both its frequency $\omega$ and the group delay at that frequency $g(\omega)$.
Thus (\ref{eqn:Z_tau}) and (\ref{eqn:Z_mu}) can be written as
\begin{eqnarray}\label{eqn:Z_tau_simple}
Z_{\tau}
&=& \frac{\frac{\partial }{\partial \tau}X_{\omega}(\tau)}{X_{\omega}(\tau)} + \beta \eta - \jmath \omega
\end{eqnarray}
and
\begin{eqnarray}\label{eqn:Z_mu_simple}
\nonumber
\lefteqn {Z_{\mu}  }\\
\nonumber
&=& \left \{ {  \frac{\frac{\partial }{\partial \omega}X_{\omega}(\tau)}{X_{\omega}(\tau)} + \frac{\mathrm{d}\beta}{\mathrm{d}\omega} \left \{  \frac{1}{\beta} + \eta(\omega) g(\omega) \right \} - \jmath g(\omega)  } \right \} \frac{\mathrm{d}\omega}{\mathrm{d}\mu}\\
& &
\end{eqnarray}
respectively, where $\eta (\omega) =  \frac{{\dot{h}^{*}} (\beta \{ g(\omega) \})}{h^{*} (\beta \{ g(\omega) \})}$. Then because $\Im \{ \eta(\omega) \} = 0$ for all three filter types \textit{including the complex gammachirp} (c.f. Appendix \ref{app:proto}), the time derivative of the phase of the integrand is
\begin{eqnarray}\label{eqn:intg_time_deriv}
\Im \{Z_{\tau} \}  &=& \Im \left \{ \frac{\frac{\partial }{\partial \tau}X_{\omega}(\tau)}{X_{\omega}(\tau)} \right \} - \omega
\end{eqnarray}
and the frequency derivative is
\begin{eqnarray}\label{eqn:intg_freq_deriv}
\Im \{ Z_{\mu} \} &=&  \left \{ \Im \left \{ \frac{\frac{\partial }{\partial \omega}X_{\omega}(\tau)}{X_{\omega}(\tau)} \right \} - g( \omega ) \right \} \frac{\mathrm{d}\omega}{\mathrm{d}\mu}
\end{eqnarray}
Time and frequency derivatives of the analysis integral (\ref{eqn:analysis}) are constructed using the derivative filters $\frac{\partial}{\partial \tau} h_{\omega} (\tau)$ and $\frac{\partial}{\partial \omega} h_{\omega} (\tau)$ respectively, c.f.  \cite{Augur1995TSP}, \cite{Kodera1978TSP}:
\begin{eqnarray}\label{eqn:dX_dtau}
  \frac{\partial }{\partial \tau}X_{\omega}(\tau) &=& x(\tau) * \frac{\partial}{\partial \tau} h_{\omega} (\tau)
\end{eqnarray}
\begin{eqnarray}\label{eqn:dX_domega}
  \frac{\partial }{\partial \omega}X_{\omega}(\tau) &=& x(\tau) * \frac{\partial}{\partial \omega} h_{\omega} (\tau)
\end{eqnarray}
Alternatively, as suggested in \cite{Kodera1978TSP}, they can be derived from the output of (\ref{eqn:analysis}) by direct differentiation of the signal $X_{\omega}(\tau)$ to obtain (\ref{eqn:dX_dtau}) and by using neighbouring analysis filters to obtain an approximation to (\ref{eqn:dX_domega}).
Stationary phase points $\{(\omega_{i}, \tau_{i})\}_{i}$ are solutions to:
\begin{eqnarray} \label{eqn:stationary_phase_points}
  \Im \{Z_{\tau} \} &=& \Im \{ Z_{\mu} \} = 0
\end{eqnarray}
The derivative filters can be expressed in terms of the prototype filter $h(t)$ and its derivative $\dot{h} (t)$.
Expressions for the derivative $\dot{h}(t)$ of the Gaussian and gammachirp filters can be found in Appendix \ref{app:proto}.
%
%
%
%
From (\ref{eqn:stationary_phase_points})
there are two conditions that must be satisfied simultaneously for a
stationary phase point to occur at $(\omega_{i}, \tau_{i})$, specifically:
\begin{enumerate}
  \item the frequency, $\frac{ \partial \angle X_{\omega}(\tau)
  }{\partial\tau} = \Im \left \{   \frac{\frac{\partial }{\partial \tau}X_{\omega}(\tau)}{X_{\omega}(\tau)}   \right \}$, observed at the output of filter at time $\tau$, is equal to the
centre frequency $\omega$ of the filter;
  \item the delay, $\frac{ \partial \angle X_{\omega}(\tau)
  }{\partial\omega}  = \Im \left \{   \frac{\frac{\partial }{\partial \omega}X_{\omega}(\tau)}{X_{\omega}(\tau)}   \right \}$, observed
at the output of the filter at frequency $\omega$, is equal to the
group delay $g(\omega)$ of the filter at that frequency.
\end{enumerate}
Together these define a \textit{signal matching condition}: at
$(\omega, \tau)$ the frequency and delay observed at the output of
the filter must match the designed centre frequency and group delay of
that filter.

Locating stationary phase point requires a grid search over $\omega$ for a bank of analogue filters or over both $\omega$ and $\tau$, for a discrete-time filter bank. Such a grid search is not onerous since it is implicit in the implementation of the analysis integral.  With a grid search there is always the risk of missing the pair $(\omega_{i}, \mu_{i} )$ that satisfy (\ref{eqn:stationary_phase_points}).
This risk can be reduced by: (i) defining a phase gradient vector
\begin{eqnarray}\label{eqn:phase_rate_vector}
    \boldsymbol{\phi} (\tau , \mu )&\triangleq& [ \Im \{Z_{\tau}  \} \,\,\, \Im \{ Z_{\mu}  \} ]^{T}
\end{eqnarray}
where the superscript $T$ indicates matrix transpose;
(ii) using the Euclidean norm of this vector to construct
a test for stationary phase points, i.e.
\begin{eqnarray}\label{eqn:stationary_phase_test}
\| \boldsymbol{\phi} (\tau , \mu ) \| & < & C_{1}
\end{eqnarray}
where the threshold $C_{1}$ is a small positive real constant. The Euclidean norm is used here for analytic convenience. Other vector norms may be appropriate and may have desirable properties with respect to ease of implementation.

In addition to finding stationary phase points, equations (\ref{eqn:Z_tau_simple}) and (\ref{eqn:Z_mu_simple}) can also be used to test for phase-rate dominance in the TF plane.
For phase-rate dominance the inequality
\begin{eqnarray}\label{eqn:phase_dominance}
   \frac {p(\tau , \mu )}{ \|  \mathbf{a} (\tau , \mu )  \|}& > &  C_{2}
\end{eqnarray}
must be satisfied, where
the amplitude gradient vector is
\begin{eqnarray}\label{eqn:amp_vector}
    \mathbf{a} (\tau , \mu )&\triangleq& [ \Re \{Z_{\tau}  \} \,\,\, \Re \{ Z_{\mu} \} ]^{T}
\end{eqnarray}
and
the threshold $C_{2}$ is a positive real constant greater than or equal to one.
The projection term
\begin{eqnarray}\label{eqn:projection}
 p(\tau , \mu ) &=& \frac{|\boldsymbol {\phi}^{T} (\tau , \mu ) \mathbf{W} \,\mathbf{a} (\tau , \mu)|}{\|\mathbf{a} (\tau , \mu) \|}
\end{eqnarray}
is formed from the sum of the projection of the phase rate vector in the direction of the normalised amplitude rate, i.e. ${\boldsymbol {\phi}^{T}  \mathbf{a} }/{\|\mathbf{a}  \|}$, plus the projection in the orthogonal direction, i.e. ${\boldsymbol {\phi}^{T}
\left [
\begin{smallmatrix}
  0 & 1 \\
  -1 & 0
\end{smallmatrix}
\right ]
\mathbf{a} }/{\| \mathbf{a}  \|}$ and hence $\mathbf{W} = \left [
\begin{smallmatrix}
  1 & 1 \\
  -1 & 1
\end{smallmatrix}
\right ]$.
The test is derived in Appendix \ref{app:PSFNAI2D}.
Thus the PSP divides the time frequency plane into two regions: a region $S_{0}$ where (\ref{eqn:phase_dominance}) is satisfied and the rest of the TF plane $\tilde{S}_{0}$ where it is not.

Given the stationary phase points  $\{( \mu_{i}, \tau_{i})\}_{i}$ that are solutions to (\ref{eqn:stationary_phase_points}), the
stationary phase approximation is invoked by replacing
(\ref{eqn:synthesis_integral}) by:
\begin{eqnarray}\label{eqn:synthesis_integral_SPA}
  \hat{x}(t) &\approx& \frac{1}{C} \Re \left \{ \int \! \!  \! \! \int_{S} \!  X_{\omega}(\tau) h_{\omega}^{*} (\tau-t) \, \mathrm{d}\mu \mathrm{d}\tau \right \}
\end{eqnarray}
where $S$ is a subset of the TF plane defined as $S = \{\bigcup_{i}S_{i}\}\cup \tilde{S}_{0}$, $S_{i}$ is the neighbourhood of the
$i$th stationary phase point such that $(\mu_{i}, \tau_{i}) \in S_{i}$ and $\bigcup_{i}S_{i}$ contains all points in the TF plane that satisfy (\ref{eqn:stationary_phase_test}).
Equation
(\ref{eqn:synthesis_integral_SPA}) promises
sparsity directly from analysis without the computationally
expensive re-synthesis step associated with most methods for sparse
atomic decomposition.
The atomic
decomposition of (\ref{eqn:synthesis_integral_SPA}) is sparse in the
sense that $\forall \, (\mu,\tau) \ni S$ the coefficient
$X_{\omega}(\tau)$ is implicitly set to zero.
However there are no guarantees about the degree of sparsity that can be achieved or the quality of reconstruction that might be expected apart from the usual ones that might be expected from a well-designed snug or tight frame \cite{strahl:2379}.
This will be explored in the following section.
Together the analysis steps of (\ref{eqn:analysis}), (\ref{eqn:Z_tau_simple}) \& (\ref{eqn:Z_mu_simple}), the selection inequalities (\ref{eqn:stationary_phase_test}) \& (\ref{eqn:phase_dominance}) and the synthesis equation (\ref{eqn:synthesis_integral_SPA}) form what might be called a time-frequency stationary phase approximation (TFSPA).

\section{Elementary Signals}
\label{sec:elementary}
This Section
is devoted to a consideration of the stationary phase regions of the TF plane
associated with some important elementary signals and also in identifying regions of phase-rate dominance.
Because of the elementary nature of the signals there is some hope that closed form solutions are possible.
These elementary waveforms are: an impulse; a single tone; a linear chirp; a decaying phasor.
Closed form solutions that define the stationary phase points are derived.
Within the constraints of the available space results of the numerical evaluation of these regions are also provided to confirm and expand upon the theoretical results.

 The results are for order $n=4$ gammatone ($c=0$) and gammachirp ($c=4$) filter banks at the cochlear spacing described in Section \ref{sec:prelim}. Numerical evaluation considers a frequency band from $\omega_{min} = 2 \pi \times 100$ rad/s to $\omega_{max} = 2 \pi \times 5000$ rad/s at a sampling rate of $20$ kHz. For the gammatone, $\frac{\mathrm{d}\omega}{\mathrm{d\mu}} \frac{1}{\beta} = 25 $ and $\frac{\mathrm{d} \beta}{\mathrm{d}\omega} = \frac{1}{9}$ and for the gammachirp, $\frac{\mathrm{d}\omega}{\mathrm{d\mu}} \frac{1}{\beta} = 36 $ and $\frac{\mathrm{d} \beta}{\mathrm{d}\omega} = \frac{1}{13}$. These values are commensurate with ERB figures for the cochlea, c.f. \cite{Moore1987209}. The spacing of the filters in frequency is such that there are 4 filters with their maximum gain within the ERB of each filter \cite{strahl:2379}. In this case 103 filters are used to cover the stated band. The combined frequency response of the analysis and synthesis filters banks has a linear phase and is flat to within a fraction of a dB over the
band. Thus, over the band of interest, the combination of analysis and synthesis  filter banks, (\ref{eqn:analysis}) \& (\ref{eqn:synthesis_integral}), is effectively distortionless. All filters are approximated by simply truncating them at a suitable point to give finite impulse response (FIR) filters. Anticausal filters (e.g. synthesis filters) are simulated by incorporating suitable delays.

For the inequalities ((\ref{eqn:stationary_phase_test}) and (\ref{eqn:phase_dominance})), $C_{1} = 10$ and $C_{2} = 1$. The former was found experimentally to provide a good indication of the stationary phase regions for the filter bank described above and for all combinations of filter spacing and prototype filter described in Section \ref{sec:prelim}. The latter is an extreme value. The inequalities, (\ref{eqn:phase_dominance}) of Section \ref{sec:PSFNAI} and (\ref{eqn:phase_rate_dominance_app}) of Appendix \ref{app:PSFNAI2D}, suggest a larger value such as $C_{2} = 10$. However a value of unity was chosen because: (i) it makes clear the boundary in the TF plane between regions where the projection $p (\tau , \mu)$  is greater than  the norm $\| \mathbf{a} (\tau , \mu) \|$; (ii) it reflects a desire to test the degree of sparsity that could be obtained.
Thus (\ref{eqn:phase_dominance}) becomes
\begin{eqnarray}\label{eqn:phase_dominance2}
  p (\tau , \mu) &>&  \| \mathbf{a} (\tau , \mu) \|
\end{eqnarray}
Studies of ridgelets and skeletons as in \cite{Carmona1999TSP} suggest that for asymptotic signals, very sparse representations are possible.

\subsection{An impulse}
The impulse is obviously important because it is an extreme example of a transient signal. Further it is not an asymptotic signal in the sense considered in \cite{Delprat1992IT} and as such illustrates the advantages of applying to PSP to the synthesis integral rather than the analysis integral.
For an impulse $x(t) = \delta (t)$ the analysis of (\ref{eqn:analysis}) yields $X_{\omega} (\tau) = \beta h(\beta\tau) \mathrm{e}^{\jmath \omega \tau}$.
From which, (\ref{eqn:Z_tau_simple}) and (\ref{eqn:Z_mu_simple}) give
\begin{eqnarray}\label{eqn:gamma_impulse_dtau}
Z_{\tau}  &=&
\beta \left \{ \frac{\dot{h}(\beta \tau)}{h(\beta \tau)} + \eta \right \}
\end{eqnarray}
and
\begin{eqnarray}\label{eqn:gamma_impulse_domega}
\nonumber
\lefteqn {Z_{\mu} =} \hspace{3.0in}\\
\nonumber
\lefteqn{  \frac{\mathrm{d}\omega}{\mathrm{d}\mu} \frac{1}{\beta}  \left\{ \frac{\mathrm{d} \beta}{\mathrm{d} \omega} \{ {2} + \beta \tau \frac{\dot{h}(\beta \tau)}{h(\beta \tau)} + \beta g(\omega) \eta \} + \jmath  \{ \beta \tau - \beta g(\omega)\}   \right \}} \hspace{3.0in}\\    \end{eqnarray}
For a Gaussian pulse $g(\omega) =0$, $\eta = 0$ and $\frac{\dot{h}(\beta \tau)}{h(\beta \tau)} = - \beta \tau $. For a gammachirp pulse $g(\omega) = \frac{n}{\beta}$, $\eta = -1/n$ and $\frac{\dot{h}(\beta \tau)}{h(\beta \tau)} = -1 + \frac{n-1}{\beta \tau} + \jmath c \left \{ \frac{1}{\beta \tau} - \frac{1}{n} \right \}$, c.f. Appendix \ref{app:proto}. Thus, for all the three pulse type at filter spacing considered, the stationary phase region is defined as $\tau = g(\omega) $, $\forall\omega$, i.e. a contour in the TF plane at the group delay.

As to phase-rate dominance, the most straightforward case to consider is
a uniform filterbank constructed from Gaussian filters for which
 $Z_{\tau} = - \beta ^{2} \tau$, $Z_{\mu} = \jmath \frac{\mathrm{d}\omega}{\mathrm{d}\mu} \tau$, $\boldsymbol{\Phi} ^{T} = \left [ 0 \,\,\, \frac{\mathrm{d}\omega}{\mathrm{d}\mu} \tau \right]$ and $\mathbf{a}^{T} = \left [  - \beta ^{2} \tau \,\,\, 0\right ]$.
Using (\ref{eqn:phase_dominance}) and (\ref{eqn:projection}), the test for phase dominance becomes
\begin{eqnarray}\label{eqn:impulse_Gauss_theory}
 p(\tau , \mu ) &=& \frac{|\boldsymbol {\phi}^{T} (\tau , \mu ) \mathbf{W} \,\mathbf{a} (\tau , \mu)|}{\|\mathbf{a} (\tau , \mu) \| ^{2}}  = \frac{1}{\beta ^{2}} \frac{\mathrm{d}\omega}{\mathrm{d}\mu} > C_{2}
\end{eqnarray}
Recalling that for uniform filter banks $\frac{\mathrm{d}\omega}{\mathrm{d}\mu} = \omega _{max} - \omega_{min}$ and thus
\begin{eqnarray}
\nonumber
   \beta & < & \sqrt{\frac{\omega _{max} - \omega_{min}}{C_{2}}}
\end{eqnarray}
In the following sub-section this result will be considered again and contrasted with a related result involving the response to a single phasor.

More generally it is worth noting that $Z_\mu$ in (\ref{eqn:gamma_impulse_domega}) is a function of normalized time $\beta \tau$ alone and is not dependent on the filter frequency $\omega$ since both $\left \{\frac{\mathrm{d}\omega}{\mathrm{d}\mu} \frac{1}{\beta} \right \}$ and $\frac{\mathrm{d} \beta}{\mathrm{d} \omega}$ are constants for a given filterbank. $Z_\tau$, on the other hand, is dependent on $\omega$ indirectly through $\beta$ if the filterbank is non-unform. If $Z_\tau$ varies with $\omega$ and $Z_\mu$ does not then the relationship between $p(\tau, \mu)$ and $|| \mathbf{a} (\tau, \mu)||$ may also be dependent on $\omega$.

For a gammachirp pulse (\ref{eqn:gamma_impulse_dtau}) and (\ref{eqn:gamma_impulse_domega}) become
\begin{eqnarray}
   Z_{\tau} &=& \beta \left \{ \frac{n-1}{\beta \tau} - \frac{n+1}{n} + \jmath c \left \{ \frac{1}{\beta \tau} -  \frac{1}{n} \right \} \right \}
\end{eqnarray}
and
\begin{eqnarray}
   Z_{\mu} &=&  \left \{\frac{\mathrm{d}\omega}{\mathrm{d}\mu} \frac{1}{\beta} \right \} \left \{ n- \beta \tau \right \} \left \{\frac{\mathrm{d} \beta}{\mathrm{d} \omega}  + \jmath \left \{ \frac{\mathrm{d} \beta}{\mathrm{d} \omega}\frac{c}{n}-1\right \}\right \}
\end{eqnarray}
respectively. Clearly both $\Re \{ Z_{\tau} \}$ and $\Im \{Z_{\mu} \} $ and hence $||\mathbf{a} (\tau,\mu) ||$ are not dependent on $c$.

\begin{figure}
  	\centering
		\includegraphics[width=0.9\linewidth]{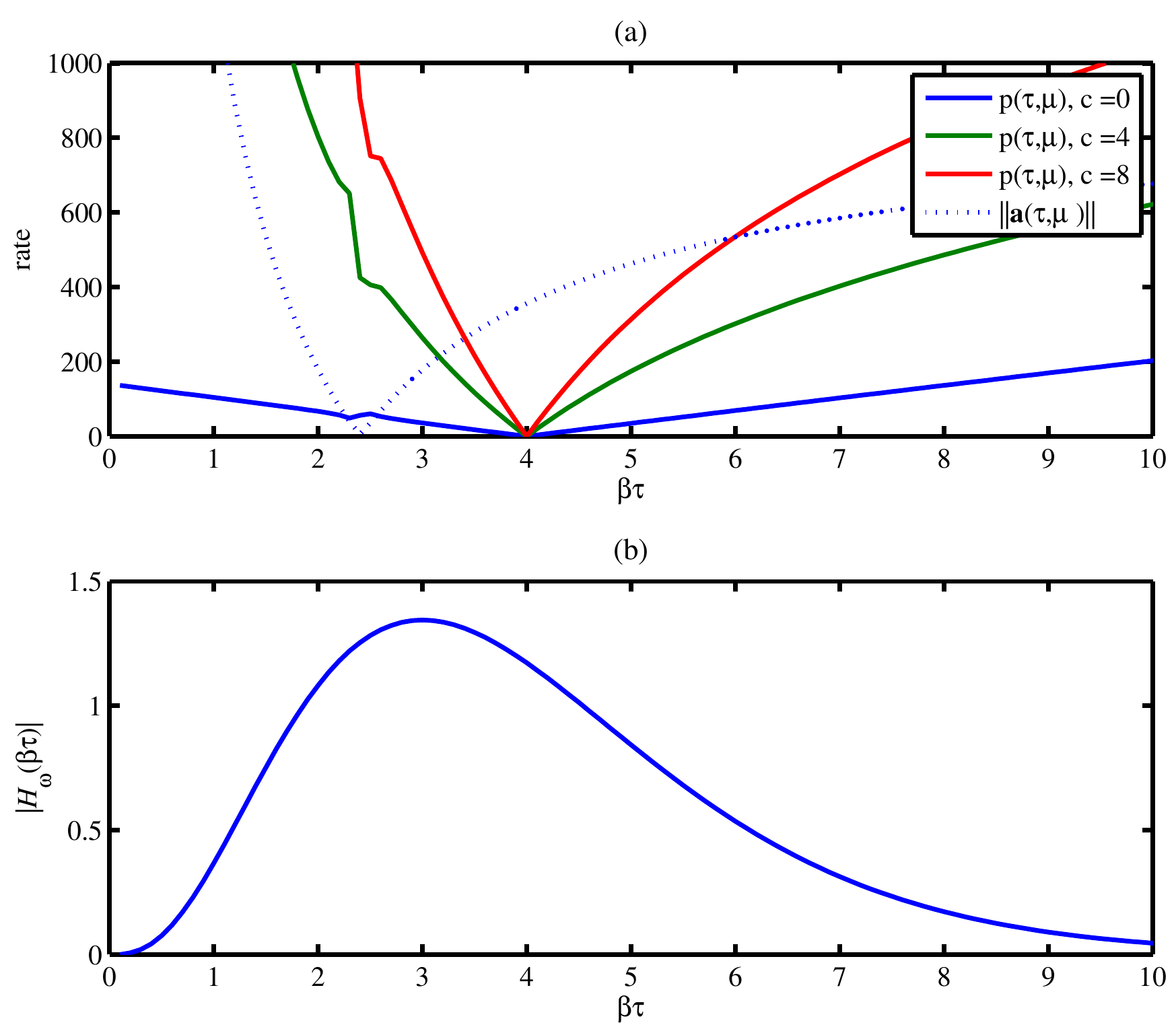}
	\caption{Impulse response of gammachirp filter at 1 kHz: (a) comparison of $p(\tau, \mu)$ and $|| \mathbf{a} (\tau, \mu)||$ at various values of chirp rate parameter $c$; (b) response.}
	\label{fig:rates_impulse_response}
\end{figure}

Figure \ref{fig:rates_impulse_response}(a)  illustrates the relationship between $p(\tau, \mu)$ and $|| \mathbf{a} (\tau, \mu)||$ for a single filter in a filterbank
where $\frac{\mathrm{d}\beta}{\mathrm{d}\omega} = \frac{1}{9}$, $\frac{\mathrm{d}\omega}{\mathrm{d}\mu} \frac{1}{\beta} = 35$ and $\beta = 712$ rad/s for 3 values of the filter parameter $c$.
The stationary phase point is at $\beta \tau = n = 4$ whereas the peak magnitude of the impulse response occurs is at $\beta \tau = n-1 = 3$, c.f. Figure \ref{fig:rates_impulse_response}(b).
Insight into the relative behaviour of  $p(\tau, \mu)$ and $|| \mathbf{a} (\tau, \mu)||$, and how that is affected by the value of $c$, can be obtained by considering either a uniform filter bank, i.e. $\frac{\mathrm{d} \beta}{\mathrm{d} \omega}=0$, or, equivalently, a non-uniform filter bank where $\frac{\mathrm{d} \beta}{\mathrm{d} \omega} \ll 1$ and $\frac{\mathrm{d} \beta}{\mathrm{d} \omega} |\frac{c}{n}| \ll 1$. The latter leads to the approximation
\begin{eqnarray}
   Z_{\mu} & \approx &  - \jmath \left \{\frac{\mathrm{d}\omega}{\mathrm{d}\mu} \frac{1}{\beta} \right \} \left \{ n- \beta \tau \right \}.
\end{eqnarray}
The amplitude rate norm is thus
\begin{eqnarray}
   || \mathbf{a} (\tau, \mu)|| &\approx& \beta \left | \frac{n-1}{\beta \tau} - \frac{n+1}{n} \right |
\end{eqnarray}
which has a zero at $\beta \tau = \{ \frac{n-1}{n+1} \} n$, c.f. Figure \ref{fig:rates_impulse_response}(a). Straightforward manipulation also gives:
\begin{eqnarray}
   p(\tau, \mu) & \approx & \left | \beta c \left \{ \frac{1}{\beta \tau} - \frac{1}{n} \right \} +  \left \{\frac{\mathrm{d}\omega}{\mathrm{d}\mu} \frac{1}{\beta} \right \} \left \{ n- \beta \tau \right \} \right |
\end{eqnarray}
which, not surprisingly, has a zero at the stationary phase point. Consider two limiting cases of (\ref{eqn:phase_dominance2}). First,  as $\beta \tau \rightarrow 0$, it reduces to
$  \left | \frac{\beta c }{\beta \tau} \right | > \beta \left | \frac{n-1}{\beta \tau} \right \|$,
which is satisfied if $|c| > n-1$. Thus, for example, in Figure \ref{fig:rates_impulse_response}(a), for $\beta\tau < 3$, the inequality is satisfied for $c \geq 4$ and phase rate dominance can be achieved to the left of the peak in the impulse response.  Second, as $\beta \tau \rightarrow \infty$ and $c \geq 0$, the inequality reduces to:
\begin{eqnarray}\label{eqn:impulse_equality_limit}
  \left \{\frac{\mathrm{d}\omega}{\mathrm{d}\mu} \frac{1}{\beta} \right \} \left \{ \beta \tau  -n \right \} &>& \frac{\beta}{n} \left \{ n+1 -c \right \}
\end{eqnarray}
The LHS is always positive and linearly increasing with $\beta \tau$, hence a value  $\beta \tau_{0} $ can always be found above which the inequality is satisfied. For $\{c: 0 \leq c < n+1 \}$, increasing $c$ reduces the RHS and hence reduces $\beta \tau_{0}$. For $c \geq n+1$ the inequality is satisfied for all $\beta \tau$. A degree of caution must be exercised in the application of this result as it is based on asymptotic arguments. However it does predict the trend. As $c$ increases from 0, $\beta \tau_{0}$ decreases, increasing the region $\{ \beta \tau : \beta \tau > \beta \tau_{0}\}$ where the inequality is satisfied. This trend is evident in Figure \ref{fig:rates_impulse_response}(a) for $\beta\tau > 4$. In summary, increasing the value of $c$ increases the region to left of the peak response where phase rate dominance can be achieved. To the right of the stationary phase point this region can also be increased by increasing and positive values of $c$. The effect of this increase will however, as for the Gaussian pulse, be influenced by the relative sizes of $\left \{\frac{\mathrm{d}\omega}{\mathrm{d}\mu} \frac{1}{\beta} \right \}$ and $\beta$, c.f. (\ref{eqn:impulse_equality_limit})

\begin{figure}
  	\centering
		\includegraphics[width=0.9\linewidth]{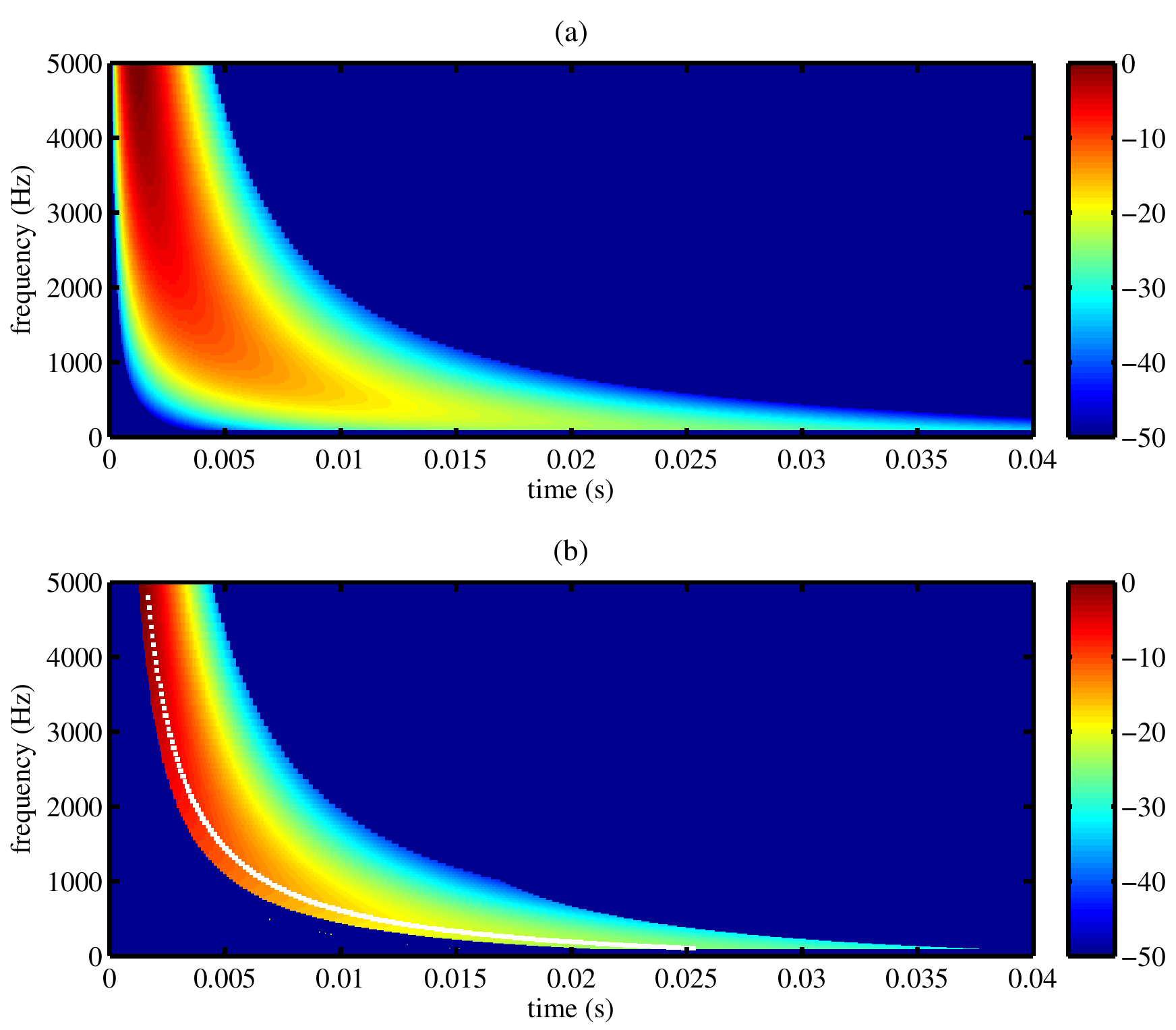}
	\caption{Response a gammachirp cochlear-spaced filter bank to single impulse: (a) $|X_{\omega} (\tau)|$ in dB; (b) TFSPA - stationary phase points in white.}
	\label{fig:gamma_impulse_tf}
\end{figure}

\begin{figure}
  	\centering
		\includegraphics[width=0.9\linewidth]{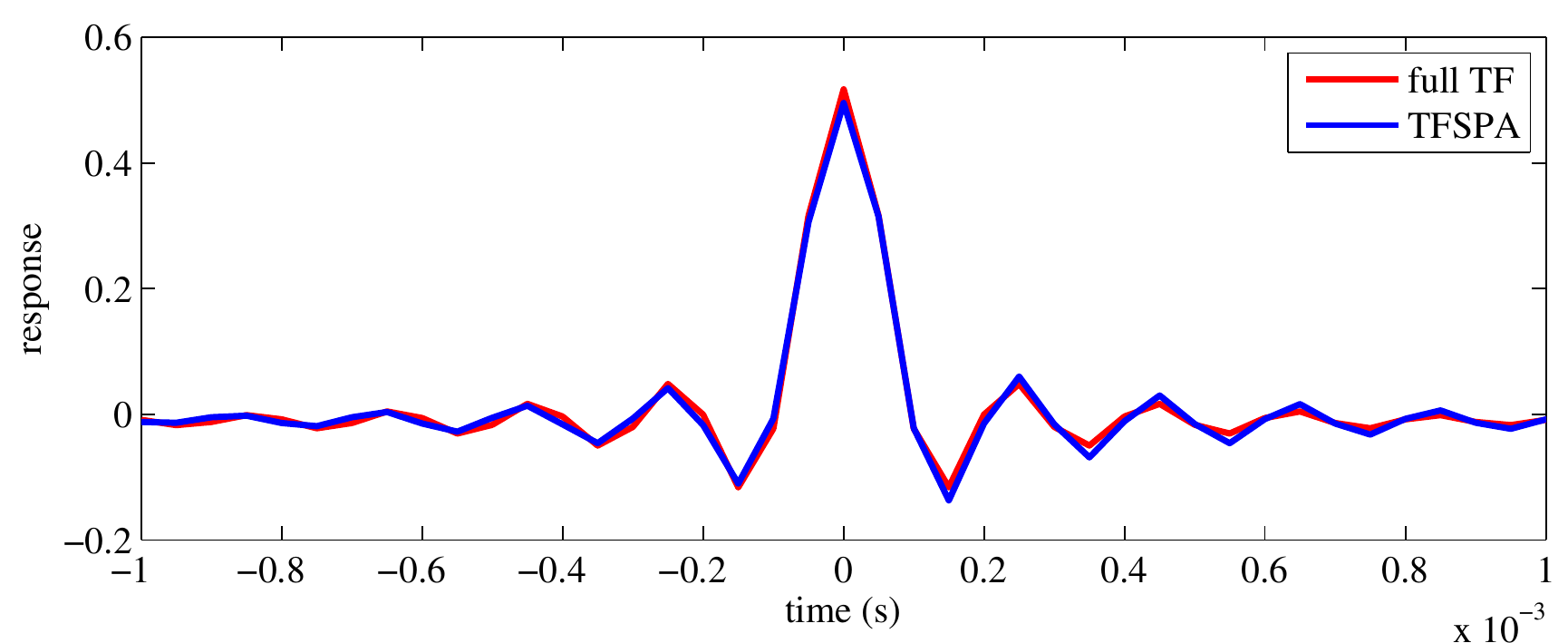}
	\caption{Reconstruction of impulse using full TF plane of Fig \ref{fig:gamma_impulse_tf}(a) and TFSPA of Fig \ref{fig:gamma_impulse_tf}(b).}
	\label{fig:gamma_impulse_recon}
\end{figure}

Figure \ref{fig:gamma_impulse_tf} shows both the complete TF response and results for TFSPA for $n=4$ and $c=4$.
The stationary phased points that satisfy (\ref{eqn:stationary_phase_test}) are indicated in white on Figure \ref{fig:gamma_impulse_tf} (b).  This stationary phase contour is not co-incident with the peak response or ridge but rather lies at the group delay of each filter as shown earlier. Figure \ref{fig:gamma_impulse_tf}(b) is obtained by only plotting $|X_{\omega} (\tau)|$ at points in the TF plane where (\ref{eqn:phase_dominance2}) is not satisfied. As expected from Figure \ref{fig:rates_impulse_response}, the leading edge of the response has been removed.
Reconstructions of the input waveform are shown in Figure \ref{fig:gamma_impulse_recon} using both full TF plane and TFSPA. In this case they are almost identical despite the removal of the leading edge of the response.

\subsection{A single phasor}
\label{sec:elementary_phasor}
Apply a single tone of $\lambda$ rad/s to the analysis filter bank i.e. $x(t) = u(t) e ^{j \lambda t}$. In the steady state for $t\gg0$ this gives
\begin{eqnarray}
\nonumber
   X_{\omega} (\tau)&\approx&  \mathrm{e} ^{\jmath \lambda \tau}  H \left( \Omega \right)
\end{eqnarray}
where $\Omega = \frac{\lambda  - \omega}{\beta}$,
with $Z_{\tau}
    = \beta \left \{\eta + \jmath \Omega \right \}$
and
\begin{eqnarray}
\nonumber
\lefteqn {Z_{\mu}= } \hspace{3.2in}\\
\nonumber
\lefteqn{
 \frac{\mathrm{d}\omega}{\mathrm{d}\mu} \frac{1}{\beta}
\left\{
\frac{d \beta}{d \omega} \{ 1 + \eta \beta g(\omega) \}
- \frac{\dot{H} (\Omega)}{H(\Omega)}  \{ 1 + \frac{d \beta}{d \omega} \Omega \}
  -\jmath \beta g(\omega)
\right\}
}\hspace{3.2in}\\
\end{eqnarray}
Thus for all filters considered at the any of the filter spacings considered the stationary phase point occurs when
$\lambda = \omega$, $\forall \tau$ since, by
definition, the group delay of the filter at $\omega$ is $g(\omega) =
- \frac{1}{\beta} \Im \left \{ \frac{\dot{H}(0)}{H(0)} \right \}$.

As to phase-rate dominance, the most straightforward case to consider is again a uniform filterbank constructed
from Gaussian filters.
For a Gaussian pulse, $g(\omega) = 0$, $\dot{H}(\Omega) = -\Omega H(\Omega)$ and $\eta = 0$. Thus
$Z_{\tau}     =  \jmath \beta \Omega$
and
\begin{eqnarray}
\nonumber
Z_{\mu}&=& \left \{\frac{\mathrm{d}\omega}{\mathrm{d}\mu} \frac{1}{\beta} \right \}      \left \{ \frac{d \beta}{d \omega} +  \Omega \left \{ 1 +  \frac{\mathrm{d}\beta}{\mathrm{d}\omega} \Omega  \right \} \right \}.
\end{eqnarray}
Noting that $Z_{\tau}$ is purely imaginary and $Z_{\mu}$ is purely real,
$|| \mathbf{a}(\tau, \mu)|| = |Z_{\mu}|$ and
$p(\tau, \mu) = \beta |\Omega|$,
 the inequality (\ref{eqn:phase_dominance}),
reduces to
\begin{eqnarray}
   \beta |\Omega| &>&  \left \{\frac{\mathrm{d}\omega}{\mathrm{d}\mu} \frac{1}{\beta} \right \}      \left | \frac{d \beta}{d \omega} +  \Omega \left \{ 1 +  \frac{\mathrm{d}\beta}{\mathrm{d}\omega} \Omega  \right \} \right |
\end{eqnarray}
For uniformly spaced filters, this  will be satisfied if $\beta > \left \{\frac{\mathrm{d}\omega}{\mathrm{d}\mu} \frac{1}{\beta} \right \} $. Note that this in direct contradiction to (\ref{eqn:impulse_Gauss_theory})
and hence it is not possible to design a Gaussian filter bank that exhibits phase-rate dominance in response to both impulse and phasor inputs.
 For non-uniform filter banks the RHS is quadratic in $\Omega$. However provided $\frac{\mathrm{d}\beta}{\mathrm{d}\omega} |\Omega | << 1$, the same general conclusion can be drawn for a large range of $\Omega$ and hence $\lambda$.

\begin{figure}
  	\centering
		\includegraphics[width=0.9\linewidth]{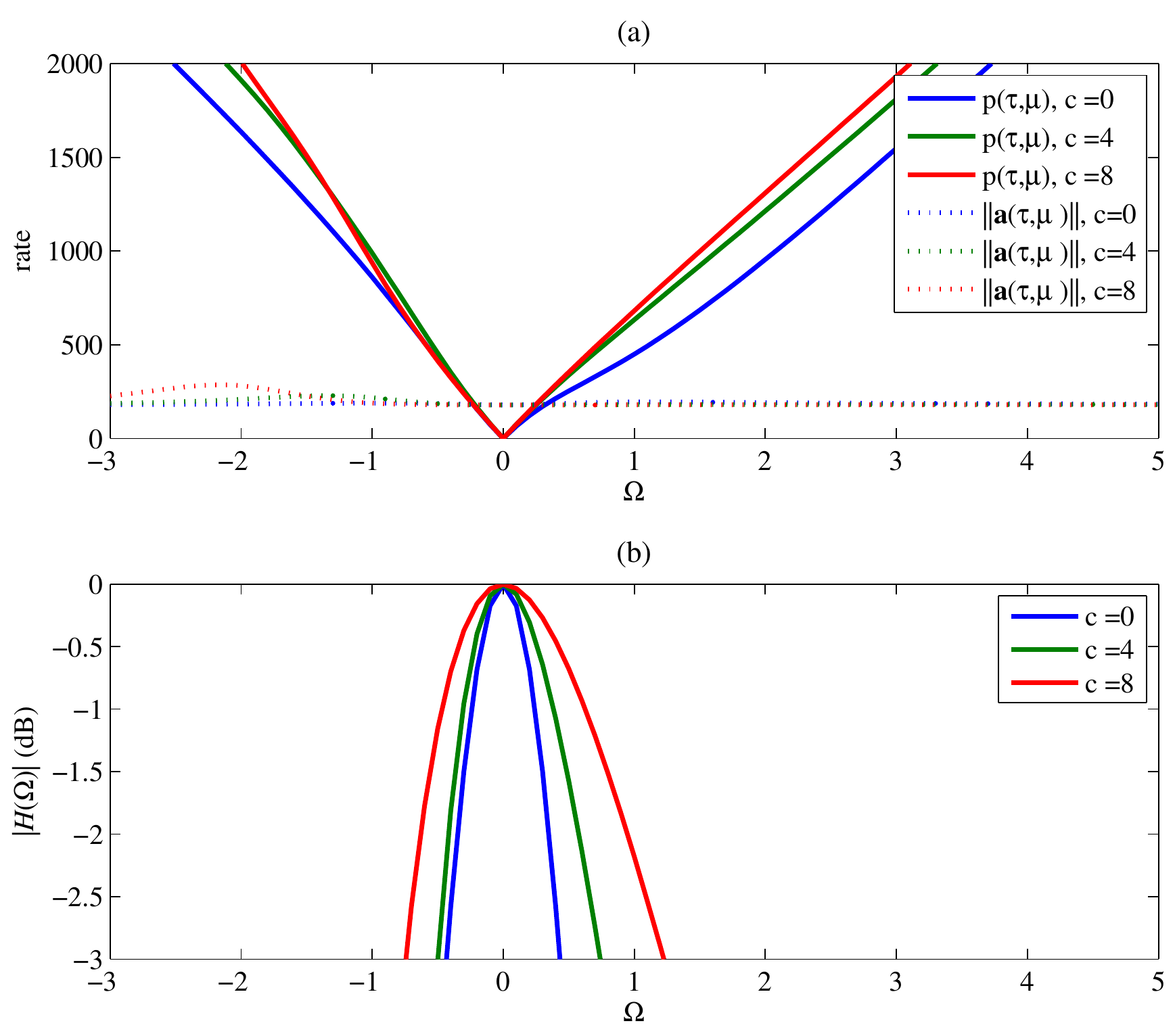}
	\caption{Steady state response of gammachirp  cochlear-spaced filter bank to 1 kHz tone.}
	\label{fig:gammachirp_freq_rate}
\end{figure}

For a \textit{gammachirp} pulse, $g(\omega) = n/\beta$, $\dot{H}(\Omega)/H(\Omega) = \frac { -n \{\Omega + j \} -\jmath c \{ \Omega +c/n \} } { 1 + \{ \Omega + c/n \}^{2} } $ and $\eta = -1/n$.
Thus
\begin{eqnarray}
\nonumber
Z_{\tau} &=& \beta \left \{ - \frac {1}{n} + \jmath \Omega \right \}
\end{eqnarray}
and
\begin{eqnarray}
\nonumber
\lefteqn {Z_{\mu} = } \\
\nonumber
    &&  \frac{\mathrm{d}\omega}{\mathrm{d}\mu} \frac{n\Omega}{\beta} \left \{    \frac {   \{ \frac{\mathrm{d}\beta}{\mathrm{d}\omega} \Omega +1  \}+ \jmath  \frac{\mathrm{d}\beta}{\mathrm{d}\omega}     +\jmath  \left \{ \Omega +\frac{c}{n} \right \} \{ \frac{c}{n} \frac{\mathrm{d}\beta}{\mathrm{d}\omega}    -    1  t \} } { 1 + \left \{ \Omega + \frac{c}{n} \right \} ^{2} }   \right \}.
\end{eqnarray}
At the stationary phase point where, $\Omega = 0$, $Z_{\tau} = -\beta/n$ and $Z_{\mu} =0$, $||\mathbf{a} (\tau, \mu) || = \beta/n$. As $|\Omega| \rightarrow \infty $,
\begin{eqnarray}
   Z_{\mu}& \rightarrow &
   \frac{\mathrm{d}\omega}{\mathrm{d}\mu} \frac{n}{\beta}
   \left \{
   \frac{\mathrm{d}\beta}{\mathrm{d}\omega}
   + \jmath \left \{
    \frac{c}{n} \frac{\mathrm{d}\beta}{\mathrm{d}\omega}    -    1
    \right \}
   \right \}
\end{eqnarray}
As in the previous sub-section, for uniform filter banks or for non-uniform filter banks where $ \frac{|c|}{n} \frac{\mathrm{d}\beta}{\mathrm{d}\omega} \ll 1$ and $ \frac{\mathrm{d}\beta}{\mathrm{d}\omega} < 1$
\begin{eqnarray}
   Z_{\mu}& \rightarrow & -\jmath \frac{\mathrm{d}\omega}{\mathrm{d}\mu} \frac{n}{\beta}
\end{eqnarray}
and hence $||\mathbf{a} (\tau, \mu) || \rightarrow \beta/n$ and
\begin{eqnarray}\label{eqn:phasor_assmp_theory}
   p(\tau, \mu)& \rightarrow & \left | \beta \Omega - \frac{\mathrm{d}\omega}{\mathrm{d}\mu} \frac{n}{\beta}  \right |.
\end{eqnarray}
Figure \ref{fig:gammachirp_freq_rate}(a)  illustrates the relationship between $p(\tau, \mu)$ and $|| \mathbf{a} (\tau, \mu)||$ for a single filter in a filterbank
where $\frac{\mathrm{d}\beta}{\mathrm{d}\omega} = \frac{1}{9}$, $\frac{\mathrm{d}\omega}{\mathrm{d}\mu} \frac{1}{\beta} = 35$ and $\beta = 712$ rad/s for 3 values of the filter parameter $c$ as in the previous sub-section. As expected from the asymptotic analysis, the norm $|| \mathbf{a} (\tau, \mu)||$ is approximately constant and virtually independent of $c$. The projection $p(\tau, \mu)$ is approximately linear in $\Omega$, as suggested by (\ref{eqn:phasor_assmp_theory}), and much less dependent on $c$ than the response to an impulse.
For these examples, (\ref{eqn:phase_dominance2}) is satisfied and phase rate dominance is achieved outside the nominal bandwidth $\beta$ of the filters, i.e. for $|\Omega| > \frac{1}{2}$. The implication is that the output of an analysis filter at $\omega$ rad/s, in response to a phasor at $\lambda$ rad/s,  contributes little to the output of the synthesis filter bank if $|\lambda - \omega| > \frac{\beta}{2}$. Further, (\ref{eqn:phase_dominance2}) can be used to locate the vicinity of the peak in the response $|X_{\omega} (\tau)|$.

\begin{figure}
  	\centering
		 \includegraphics[width=0.9\linewidth]{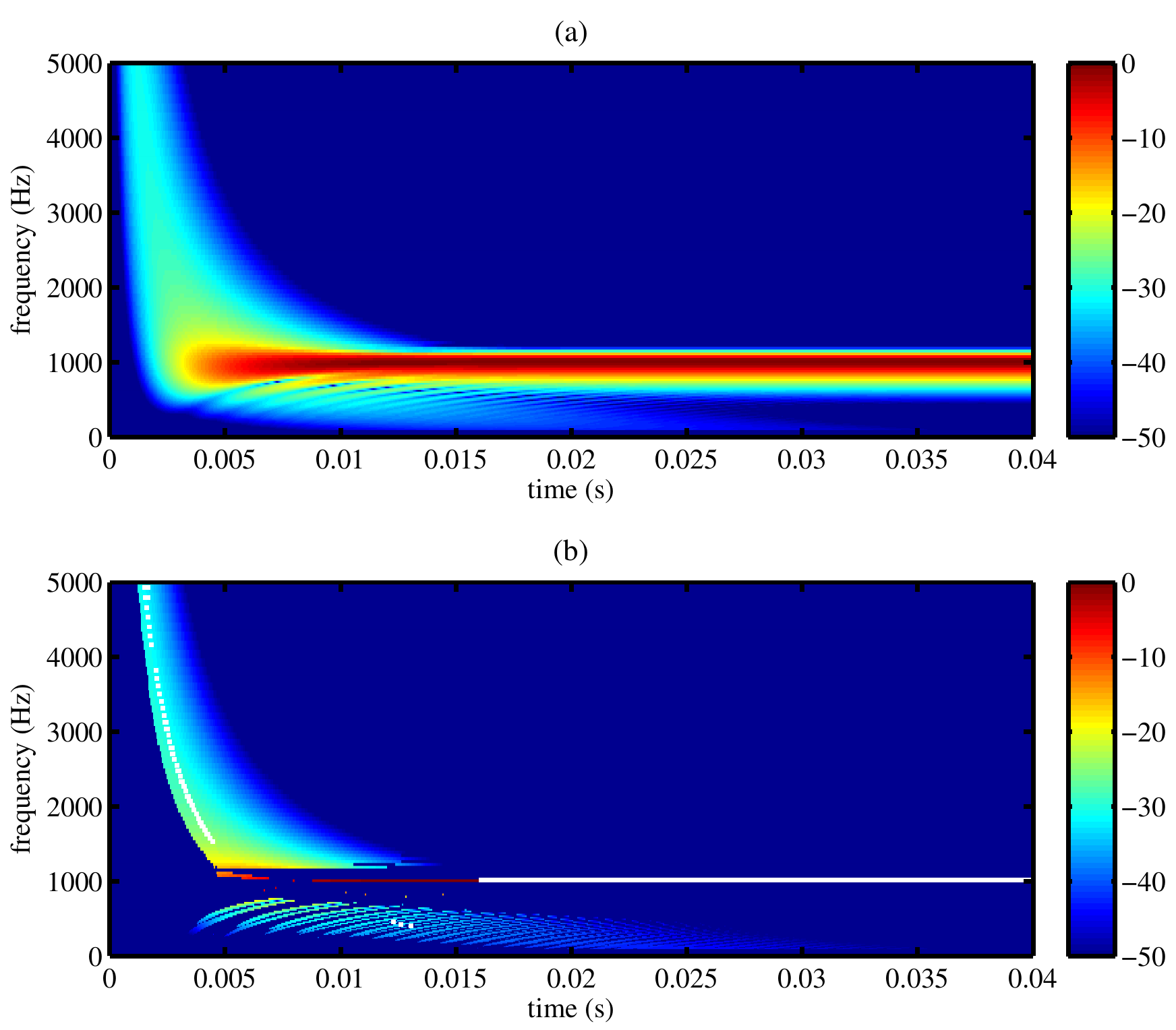}
	\caption{TF response of a gammachirp cochlear-spaced filter bank to a 1 kHz tone applied at time zero: (a) $|X_{\omega}(\tau)|$ in dB; (b) TFSPA - stationary phase points in white.}
	\label{fig:tone_gamma_tf_oct2011}
\end{figure}

\begin{figure}
  	\centering
		\includegraphics[width=0.9\linewidth]{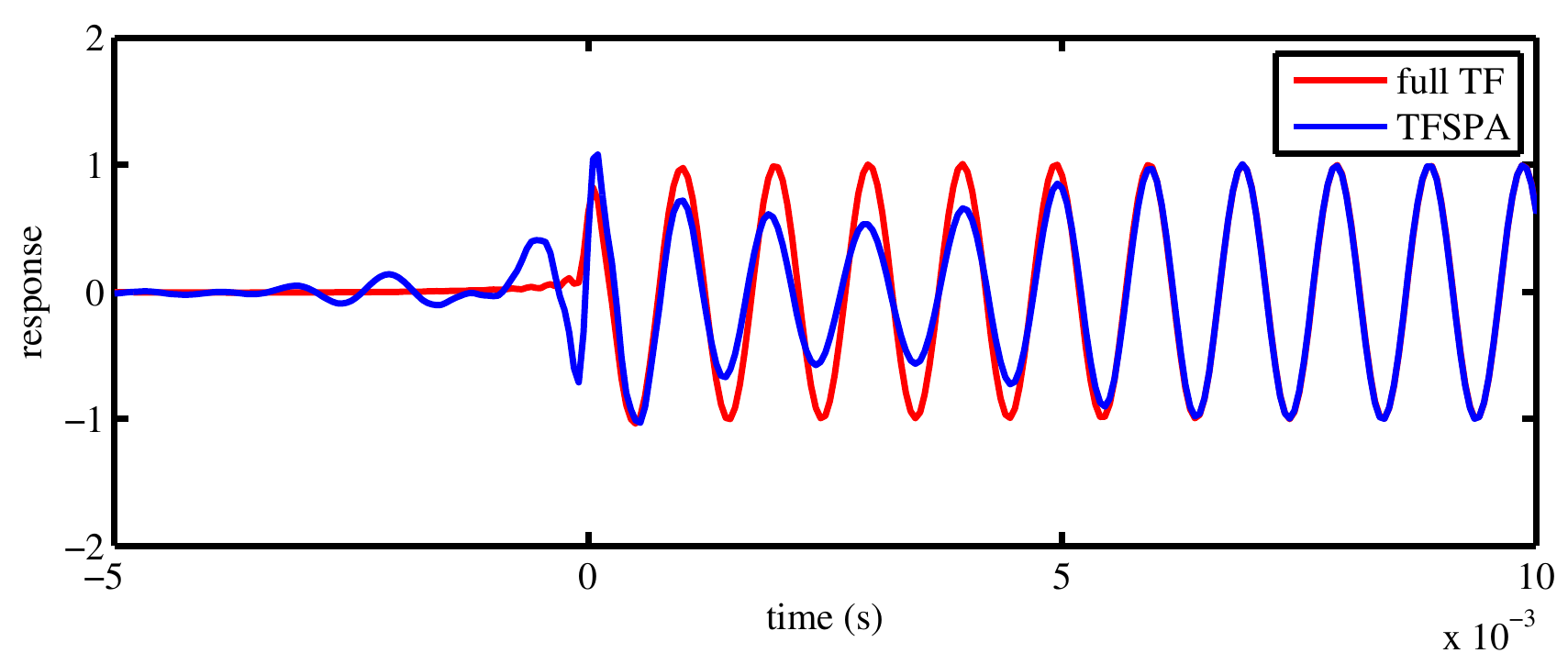}
	\caption{Reconstruction of 1 kHz tone applied at time zero}
	\label{fig:tone_gamma_recon_oct2011}
\end{figure}

Figure \ref{fig:tone_gamma_tf_oct2011} shows both the complete TF response and the results for TFSPA. The stationary phase points are indicated in white on Figure \ref{fig:tone_gamma_tf_oct2011}(b). The remaining values of $|X_{\omega} (\tau)|$ that are plotted are at points in the TF plane where (\ref{eqn:phase_dominance2}) is not satisfied. The steady state behaviour is well predicted from the theoretical considerations above as illustrated in Figure \ref{fig:gammachirp_freq_rate}. The horizontal white line at 1000 Hz corresponds with the stationary phase point on Figure \ref{fig:gammachirp_freq_rate}. There is only a small region around $\Omega = 0$ in Figure \ref{fig:gammachirp_freq_rate} where (\ref{eqn:phase_dominance2})is not satisfied. This region is just visible on Fig. \ref{fig:tone_gamma_tf_oct2011} as a horizontal band in red at 1000 Hz. The transient behaviour has similarities to that of the impulse response of Figure 2(b). Specifically the stationary phase points, on Figure 2(b) at the the group delay, are present at most frequencies on the LHS of Figure   \ref{fig:tone_gamma_tf_oct2011}(b) apart from a band of frequencies around the applied one.
Figure \ref{fig:tone_gamma_recon_oct2011} shows the reconstruction achieved both with the whole TF plane and with TFSPA. The transient response of TFSPA  introduces an amplitude modulation that dies away before matching the steady state response. This modulation is evident for $\tau < 7$ ms, the group delay of the filter, after which the stationary phase points at 1000 Hz is established.

\subsection{Linear chirp}
Consider a  linear chirp with a rate $\gamma$ rad/$\mathrm{s}^{2}$, i.e. $x(t) = u(t) \mathrm{e}^ {\jmath \frac{\gamma}{2} t ^2}$, where $u(t)$ is the step function at the origin.
The response is:
\begin{eqnarray}
\nonumber
 X_{\omega} (\tau)
 &=& \int_{0}^{\tau} \beta h(\beta t) \mathrm{e}^{\jmath \omega t} \mathrm{e}^ {\jmath \frac{\gamma}{2} \{ \tau - t\}^{2}} dt \\
 \nonumber
 &=& \mathrm{e}^{\jmath \frac{\gamma}{2} \tau^{2}} \int_{0}^{\infty} \left \{ \beta h(\beta t) \mathrm{e}^{-\jmath \gamma \tau  t} \mathrm{e}^{\jmath \frac{\gamma}{2} t^{2}} \right \} \mathrm{e}^{\jmath \omega t} dt
\end{eqnarray}
for $\tau \gg \frac{1}{\beta}$. This integral is intractable and is often approximated using the PSP. However it might appear imprudent to proceed with repeated application of that principle. Rather the approach here is to approximate the integral by more explicit means.
First assume that the frequency $\gamma t$ of the quadratic phase term $\mathrm{e}^{\jmath \frac{\gamma}{2} t^{2}}$ is approximately constant over the temporal extent of  $h(\beta t)$, specifically $\gamma \ll \beta ^{2}$.  Then replace the quadratic phase term with its frequency at the filter group delay, i.e. $\mathrm{e}^{\jmath \frac{\gamma}{2} t^{2}} \thickapprox \mathrm{e}^{\jmath \gamma g(\omega) t} $, to give
\begin{eqnarray}
\nonumber
 X_{\omega} (\tau)
&=& \mathrm{e}^{\jmath \frac{\gamma}{2} \tau^{2}} H \left ( \frac{-\omega + \gamma \tau - \gamma g(\omega)}{\beta} \right )
\end{eqnarray}
The envelope $|X_{\omega} (\tau)|$ will have a peak when $-\omega + \gamma \tau - \gamma g(\omega) = 0$, thus
$  \tau = \frac{\omega}{\gamma} + g(\omega)$.
The stationary phase points are defined as the solution to $\Im \{Z_{\tau}\}=0$, hence
\begin{eqnarray}
\nonumber
\Im \left \{ \frac{\dot{H} \left ( \frac{-\omega + \gamma \tau - \gamma g(\omega)}{\beta}\right ) }{H \left ( \frac{-\omega + \gamma \tau - \gamma g(\omega)}{\beta}\right )} \right \}
 \left \{ \frac{-1- \gamma \dot {g} (\omega) }{\beta} \right \} &=& g(\omega)
\end{eqnarray}
and $\Im \{ Z_{\mu}\} = 0$, hence
\begin{eqnarray}
   \gamma \tau  +  \Im \left \{ \frac{\dot{H} \left ( \frac{-\omega + \gamma \tau - \gamma g(\omega)}{\beta}\right ) }{H \left ( \frac{-\omega + \gamma \tau - \gamma g(\omega)}{\beta}\right )} \right \} \frac{\gamma}{\beta} &=& \omega
\end{eqnarray}
Substitution for $\Im \{ \dot{H} (.) / H(.) \}$
gives the  stationary phase points as
\begin{eqnarray} \label{eqn:chirp_stationary_points}
   \tau &=& \frac{\omega}{\gamma} +  \frac{ g(\omega)}{1+ \gamma \dot{g} (\omega)}
\end{eqnarray}
Note that $\dot{g} (\omega)$ is generally negative because $g(\omega) \propto \frac{1}{\beta}$ and $\beta$ decreases with $\omega$ (in non-uniform filter banks).
More precisely, for a Gaussian pulse $g(\omega) = 0$ and hence so is $\dot{g} (\omega)$. For log and cochlear spaced filter banks using gammatone and gammachirp pulses
$g(\omega) = \frac{n}{\beta}$ and hence $\dot{g}(\omega) = - \frac{n}{  \beta ^{2}} \frac{\mathrm{d}\beta}{\mathrm{d}\omega}  $, where $\frac{\mathrm{d}\beta}{\mathrm{d}\omega}$ is positive and $ \frac{\mathrm{d}\beta}{\mathrm{d}\omega} \ll 1$. Thus, for $\gamma \ll \beta ^ {2}$, the denominator term $1+ \gamma \dot{g} (\omega)$ lies in the interval $[0, 1)$ and thus the stationary phase points lag behind the peak response.
When the chirp rate is low, i.e. provided $\gamma \ll \beta ^{2}$,
the behaviour of the $p(\tau,\mu)$ and $\| \mathbf{a} (\tau, \mu) \|$ can be inferred from Figures \ref{fig:gammachirp_freq_rate}.   There will be a band in the TF around the stationary phase line where the (\ref{eqn:phase_dominance2}) is not satisfied.

\begin{figure}
  	\centering
		\includegraphics[width=0.9\linewidth]{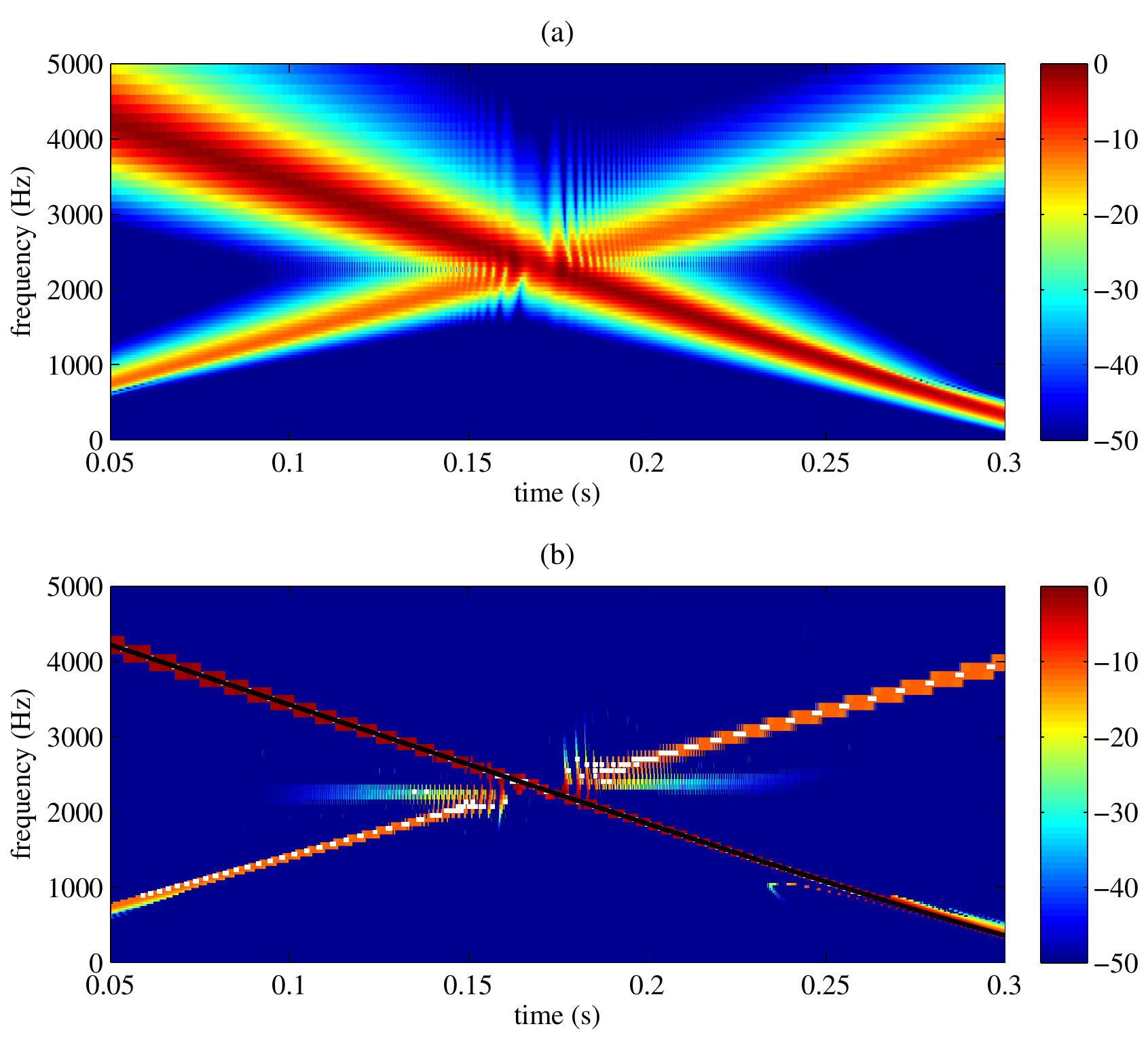}
	\caption{TF response of a gammatone cochlear-spaced filter bank to two linear chirps (a) $|X_{\omega}(\tau)|$ in dB; (b) TFSPA - stationary phase points in white - equation (\ref{eqn:chirp_stationary_points}) for downchirp in black.}
	\label{fig:chirps_oct2011}
\end{figure}

\begin{figure}
  	\centering
		\includegraphics[width=0.9\linewidth]{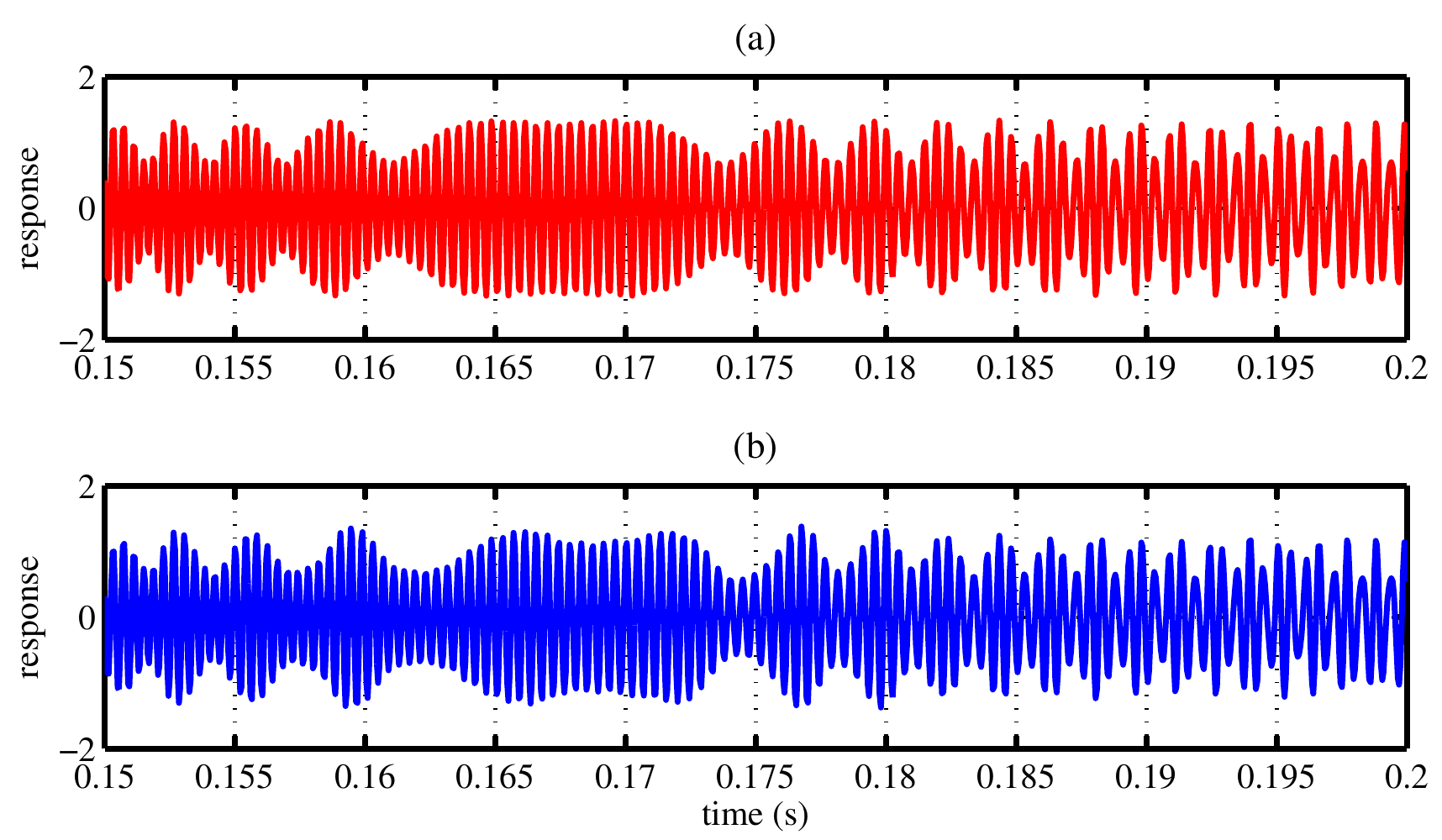}
	\caption{Reconstruction of chirps in the cross-over region using: (a) full TF plane of Fig. \ref{fig:chirps_oct2011}(a); (b) TFSPA of \ref{fig:chirps_oct2011}(b). }
	\label{fig:chirps_recon}
\end{figure}

Figure \ref{fig:chirps_oct2011} illustrates these points. The applied signal contains a down-chirp ($\gamma = -10 \times 10^4$ rad/s/s) which can be observed in Figure \ref{fig:chirps_oct2011}(a) as a continuous ridge from top left to bottom right. A second lower amplitude up-chirp ($\gamma = 8 \times 10^4$ rad/s/s) produces a ridge from bottom left to upper right. The down chirp produces a series of stationary phase points (in white) from top left to bottom right. The position of these points is well predicted by  (\ref{eqn:chirp_stationary_points}), shown as a black line, apart from bottom right where $\gamma > \beta ^2$. As expected there is a region of the TF plane around the stationary phase lines where (\ref{eqn:phase_dominance2}) in not satisfied.
The lower amplitude up-chirp is depicted in a similar manner, including stationary phase points, apart from the region where the two chirps cross. In this region the larger amplitude chirps hides the trajectory of the lower amplitude chirp. This is a form of simultaneous masking that will be explored more thoroughly in Section \ref{sec:simultaneous}.
Finally there is a low amplitude artifact that lies between the two chirps. This artifact does not contain stationary phase points and could be ignored on that basis in applications where signal analysis was the primary objective.
The outputs from the synthesis filter banks are shown in Figure \ref{fig:chirps_recon} at the point where the two frequencies cross. TFSPA produces some distortion in the reconstruction when compared with that produced using the whole TF plane.

\subsection{A decaying phasor}
Voiced speech can be viewed as a signal of the form $\sum_{\lambda_{i}} \sum_{n} A_{i} u(t-nT) \mathrm{e}^{\lambda_{i} \{ t- nT\}}$ where $T$ is the pitch period and $\{\lambda_{i}\}_{i}$ are the complex formant frequencies with centre frequencies $\{\Im (\lambda_{i}) \}_{i}$ and time constants $\{-\Re (\lambda_{i}) \}_{i}$; ${A_{i}}_{i}$ are the complex amplitudes of this atomic decomposition. Because of the relationship to speech it is informative to consider the stationary phase points associated with a single atom of this decomposition, i.e. when $x(t) = u(t)\mathrm{e}^{\lambda  t}$ with a single complex frequency $\lambda$, $\Re \{ \} < 0$. Given this:
\begin{eqnarray}
\nonumber
   X_{\omega} (\tau)
   &=& \beta  \mathrm{e}^{\lambda \tau}\int_{- \infty}^{\tau}  h(\beta t) \mathrm{e}^{\jmath \omega t} \mathrm {e} ^ {- \lambda  t} \mathrm{d}t
\end{eqnarray}
To proceed further consider the case where $h(t) \in \mathbb{R}$ and thus restrict consideration to Gaussian and gammatone pulses. The normalised time derivative is
\begin{eqnarray}
  \nonumber
  \frac{ \frac{\partial}{\partial\tau}  X_{\omega} (\tau) }{ X_{\omega} (\tau)}
   &=&  \frac { h(\beta \tau) \mathrm{e}^{\jmath \omega \tau} \mathrm {e} ^ {- \lambda  \tau}}{ \int_{- \infty}^{\tau}  h(\beta t) \mathrm{e}^{\jmath \omega t} \mathrm {e} ^ {- \lambda  t} \mathrm{d}t }   + \lambda .
\end{eqnarray}
Note that since $h(t)$ is real, the first term on the right hand side is also real if $\Im(\lambda) = \omega$ at which point: $\Im \{ Z_{\tau}\} = 0$, c.f. (\ref{eqn:Z_tau_simple}).
In a similar manner
\begin{eqnarray}
\nonumber
  \frac{ \frac{\partial}{\partial\omega}  X_{\omega} (\tau) }{ X_{\omega} (\tau)}
   &=&  \frac{\jmath \int_{- \infty}^{\tau}  h(\beta t) \mathrm{e}^{\jmath \omega t}  \mathrm {e} ^ {- \lambda  t} t \mathrm{d}t  }{\int_{- \infty}^{\tau}  h(\beta t) \mathrm{e}^{\jmath \omega t} \mathrm {e} ^ {- \lambda  t} \mathrm{d}t  }
\end{eqnarray}
which is purely imaginary for Gaussian and gammatone pulses.
Thus setting $\Im \{ Z_{\mu}\} = 0$ for $\Im \{\lambda \} = \omega $
gives
%
\begin{eqnarray} \label{eqn:decay_exp_stat_phase}
  \int_{- \infty}^{\tau}  h(\beta t) \mathrm {e} ^ {- \Re (\lambda)  t} \{t - g(\omega) \}\mathrm{d}t  &=&  0,
\end{eqnarray}
c.f. (\ref{eqn:Z_mu_simple}).
While it is unlikely that closed form solutions of (\ref{eqn:decay_exp_stat_phase}) for $\tau$ can be found, nevertheless it is still possible to infer something about the nature of the solutions. Because $h(\beta t)\mathrm {e} ^ {- \Re (\lambda)  t} > 0, \forall t$ and $\{t - g(\omega) \} < 0, \forall t < g(\omega)$, no solution exists for $\tau < g(\omega) $. For $\tau \geq g(\omega)$ the integral can be split into two terms, i.e. $ \int_{- \infty}^{\tau} \mathrm{d} t = \int_{- \infty}^{g(\omega)}\mathrm{d} t + \int_{g(\omega)}^{\tau} \mathrm{d} t$.
The first term is a negative constant. The second is is nonnegative monotonically increasing function of $\tau$ because $h(\beta t) \mathrm {e} ^ {- \Re (\lambda)  t} \{t - g(\omega) \} \geq 0, \forall \tau \geq g(\omega)$. Thus, if a solution exists it will be the only solution. Such a solution will obviously be dependent on the value of the bandwidth $\beta$ of the analysis filter for which $\omega = \Im (\lambda)$ and the time constant $-\Re (\lambda)$ of the applied pulse. Thus it may be possible to infer both the frequency and time constant of the applied pulse from the position of the stationary phase point.


\section{Simultaneous Masking}
\label{sec:simultaneous}
Simultaneous audio masking is the well studied process whereby the presence of one tone prevents the detection of a second tone \cite{MooreBook2003}.
In this section it is shown that the use of the test (\ref{eqn:stationary_phase_test}) leads to a form of simultaneous masking.
In particular, if the test (\ref{eqn:stationary_phase_test}) is used to indicate the presence or otherwise of a tonal component at $\omega$,
the amount by which that test is de-sensitized by the presence of a second tone at $\omega_{1}$ is dependent on: (i) the frequency separation of the tones $\{ \omega_{1} - \omega \}$ ; (ii) the relative amplitude $|A_{1}|$ of the tones; (iii) the magnitude frequency response $ \left | H \left ( \frac{\{ \omega_{1} - \omega \}}{\beta} \right ) \right | $ of the filter at $\omega$.
Consider a signal $ x(t) =  \mathrm{e}^ {\jmath \omega t} + A_{1} \mathrm{e}^ {\jmath \omega_{1} t}$ made up of two tones at $\omega$ and $\omega_{1}$,
where $A$, the relative amplitude, is positive real.
The steady-state response of the analysis filter at $\omega$ is
\begin{eqnarray}
\nonumber
   X _{\omega} (\tau)
   &=& \mathrm{e}^ {\jmath \omega \tau} +  A_{1} H_{1}  \mathrm{e}^ {\jmath \omega_{1} \tau}
\end{eqnarray}
where $H_{1} = H(\Omega_{1})$ and $\Omega_{1} = \{\omega_{1} - \omega \}/\beta$  is the normalized frequency separation with derivative
$\frac{\mathrm{d}\Omega_{1}}{\mathrm{d}\omega} = -
\frac{1}{\beta} \left \{ \Omega_{1}\frac{ \mathrm{d} \beta }{
\mathrm{d}   \omega} +1 \right \}$.
The time derivative follows:
\begin{eqnarray}
\nonumber
  \frac{\partial}{\partial \tau}  X_{\omega} (\tau)
  &=&  \mathrm{e}^ {\jmath \omega \tau} \jmath \omega
  +  A_{1} H _{1} \mathrm{e}^ {\jmath \omega_{1} \tau} \jmath \omega_{1}
\end{eqnarray}
as does the frequency derivative:
\begin{eqnarray}
\nonumber
    \frac{\partial}{\partial \omega }  X_{\omega} (\tau)
    &=& - \frac{1}{\beta} \dot{H}_{0}  \mathrm{e}^ {\jmath \omega \tau}
    + {A_{1}} \frac{\mathrm{d}\Omega_{1}}{\mathrm{d}\omega}
     \dot{H}_{1}  \mathrm{e}^ {\jmath \omega_{1} \tau}
\end{eqnarray}
It is convenient to
rewrite the time derivative (\ref{eqn:intg_time_deriv}) of the phase of the integrand as
\begin{eqnarray}\label{eqn:time_deriv_ratio}
\Im \{Z_{\tau}  \} &=& \frac { \Im \left \{ X_{\omega}^{*}
(\tau) \frac{\partial }{\partial \tau}X_{\omega}(\tau) \right \} -
\omega |X_{\omega}(\tau)| ^ {2} }{|X_{\omega}(\tau)| ^ {2}}
\end{eqnarray}
and the frequency derivative (\ref{eqn:intg_freq_deriv}) as
\begin{eqnarray}\label{eqn:freq_deriv_ratio}
\nonumber
\Im \{ Z_{\mu} \} &=& \left \{ {  \frac { \Im \left \{
X_{\omega}^{*} (\tau) \frac{\partial }{\partial
\omega}X_{\omega}(\tau) \right \} - g( \omega ) |X_{\omega}(\tau)| ^
{2} }{|X_{\omega}(\tau)| ^ {2}} } \right \} \frac{d\omega}{d\mu}\\
\end{eqnarray}
with denominator
\begin{eqnarray}\label{eqn:magnitude}
   | X_{\omega} (\tau) |^{2}&=&
   1  + |A_{1} H_{1}|^{2}  + 2   |A_{1} H_{1}| \cos (\theta)
\end{eqnarray}
and numerator of (\ref{eqn:time_deriv_ratio})
\begin{eqnarray}\label{eqn:norm_time_derivative}
\nonumber
\lefteqn{ \Im \left \{ X_{\omega}^{*} (\tau)  \frac{\partial}{\partial \tau}  X_{\omega} (\tau) \right \} - \omega | X_{\omega} (\tau) |^{2}
= }\\
& & |A_{1} H_{1}| \beta \Omega_{1} \left \{ |A_{1} H_{1}| + \cos
(\theta) \right \}.
\end{eqnarray}
where $\theta = \{ \omega - \omega_{1} \} \tau - \phi$ and $\phi = \angle \{A_{1} H_{1} \}$.
The angle variable $\theta$ indicates the position in time with respect to one period of the beat frequency $\{ \omega - \omega_{1} \}$.
In (\ref{eqn:freq_deriv_ratio})
\begin{eqnarray} \label{eqn:norm_freq_deriv_signal_part}
\nonumber
\lefteqn{  \Im \{X_{\omega}^{*} (\tau)  \frac{\partial}{\partial \omega}  X_{\omega} (\tau) \}
       = } \\
\nonumber
       & & \Im \left \{ -\frac{1}{\beta} \dot{H}_{0}
       + \frac{\mathrm{d}\Omega_{1}}{\mathrm{d}\omega} |A_{1}|^{2} H_{1}^{*} \dot{H}_{1}
        -\frac{1}{\beta} A_{1}^{*} H_{1}^{*} \dot{H}_{0} \mathrm{e} ^ {\jmath \{ \omega - \omega_{1} \} \tau} \right \}  \\
       & &  + \Im \left \{ \frac{\mathrm{d}\Omega_{1}}{\mathrm{d}\omega}
       A_{1} \dot{H}_{1} \mathrm{e} ^ {\jmath \{ \omega_{1} - \omega \} \tau} \right \}
\end{eqnarray}
which, unlike the numerator in (\ref{eqn:time_deriv_ratio}), is dependent on the particular pulse used.

Consider a \textit{gammatone} pulse for which $g(\omega) =
-n/\beta$, $H(\Omega ) = \frac{1}{\{ 1 + \jmath \Omega \} ^ {n}}$
and $\frac{ \dot{H} (\Omega)}{H (\Omega)} =  \frac{-n \{ \Omega + j
\}}{1 + \Omega ^ {2}}$. Hence $\Im \{ \dot{H}_{0} \}=  -n$, $\Im \{
H_{1}^{*} \dot{H}_{1}\} =  \frac{-n |H_{1}|^{2} }{1 +
\Omega_{1}^{2}}$ and $   H_{1}^{*} \dot{H}_{0}
    =    - \jmath n  H_{1}^{*}$
from which the normalized frequency derivative is obtained using (\ref{eqn:norm_freq_deriv_signal_part}).
\begin{eqnarray}
\nonumber
\lefteqn{  \Im \{ X_{\omega}^{*}(\tau) \frac{\partial}{\partial \omega} X_{\omega}(\tau) \} - g(\omega) |X_{\omega} (\tau)|^{2} = }\\
\nonumber
  & &   -\frac{n}{\beta} \frac{  \Omega_{1}  }{1 + \Omega_{1} ^ {2}} | A_{1} H_{1}
  |\{ \Omega_{1} - \frac{\mathrm{d}\beta}{\mathrm{d}\omega} \} \{|A_{1}H_{1}| + \cos(\theta)
  \}\\
  & &   -\frac{n}{\beta} \frac{  \Omega_{1}  }{1 + \Omega_{1} ^ {2}} | A_{1} H_{1}
  | \left \{ \Omega_{1}\frac{\mathrm{d}\beta}{\mathrm{d}\omega} + 1 \right \}  \sin (\theta)
  \label{eqn:norm_freq_derivative}
 \end{eqnarray}

\begin{figure*}[!t]
\normalsize
\setcounter{MYtempeqncnt}{\value{equation}}
\setcounter{equation}{51}
\begin{eqnarray}\label{eqn:gamma_norm_min_monty2}
\nonumber
\| \mathbf{\Phi} (\tau, \mu) \|^{2}
   &  \propto &   \frac {   \beta^{2}  \left \{ |A_{1} H_{1}| + \cos
(\theta) \right \}^{2} + \left \{ \frac{d \omega}{d \mu}
 \frac {n} {\beta}
 \right \}^ {2}  \frac{1 + \left \{\frac{\mathrm{d}\beta}{\mathrm{d}\omega} \right \}^2}{1 + \Omega_{1}^{2}}     \left \{  \sin (\psi)\left \{ |A_{1} H_{1}| + \cos
(\theta) \right \} +\cos(\psi)\sin(\theta) \right \} ^{2} }
 { \left \{ 1  + |A_{1} H_{1}|^{2}  + 2   |A_{1} H_{1}| \cos (\theta)\right \}^{2}}\\
\end{eqnarray}
\setcounter{equation}{\value{MYtempeqncnt}}
\hrulefill
\vspace*{4pt}
\end{figure*}
After some manipulation this yields equation (\ref{eqn:gamma_norm_min_monty2}) (see over),
where
\begin{eqnarray}
   \sin (\psi) &=&  \frac{  \Omega_{1} - \frac{\mathrm{d}\beta}{\mathrm{d}\omega} }{\sqrt{1+ \{ \frac{\mathrm{d}\beta}{\mathrm{d}\omega} \}^{2}} \sqrt{1 + \Omega_{1}^{2}}}
\end{eqnarray}
\begin{eqnarray}
   \cos (\psi) &=&  \frac{  \Omega_{1} \frac{\mathrm{d}\beta}{\mathrm{d}\omega} +1 }{\sqrt{1+ \{ \frac{\mathrm{d}\beta}{\mathrm{d}\omega} \}^{2}} \sqrt{1 + \Omega_{1}^{2}}}.
\end{eqnarray}
The numerator of (\ref{eqn:gamma_norm_min_monty2}) is the sum of two oscillatory terms in $\theta$.
Assuming that the filterbank is such that $\beta ^2 \gg  \left \{ \frac{d \omega}{d \mu}
 \frac {n} {\beta}
 \right \}^ {2} \left \{ 1 + \left \{\frac{\mathrm{d}\beta}{\mathrm{d}\omega} \right \}^2 \right \}$,
then the first oscillatory term, from $\Im \{ Z_{\tau}\}$, is larger than the second term, from $\Im \{ Z_{\mu}\}$.
To obtain an approximation to the norm, first consider
$|A_{1}H_{1}| \ll 1$,
in which case the denominator of (\ref{eqn:gamma_norm_min_monty2}) is a constant.
For $|\Omega_{1}| \gg 1$, $|\sin (\psi)|\approx 1$,  $|\cos(\psi)| \approx \frac{\mathrm{d}\beta}{\mathrm{d}\omega}$ (i.e. small and finite) and $\frac{\mathrm{d}\beta}{\mathrm{d}\omega} \approx 0$. Thus the numerator is the sum of two oscillatory terms proportional to $ \left \{ |A_{1} H_{1}| + \cos
(\theta) \right \}^{2}$.
Thus the minima of the norm  are, approximately, at $\cos (\theta) = - A_{1}H_{1}$.
For $\Omega_{1} \ll 1$, $|\cos(\psi)| \approx 1$ and $|\sin(\psi)| \approx 0$ and hence the two oscillatory terms are approximately out of phase and thus the minima of the numerator can again be found at the zeros of the larger term, i.e. $\cos (\theta) = - A_{1}H_{1}$.
Substitution in (\ref{eqn:gamma_norm_min_monty2}) yields
\begin{eqnarray}\label{eqn:approx_gamma}
\|  \mathbf{\Phi} (\tau , \mu ) \| _{\mathrm{min}} &\approx& \frac{d \omega}{d \mu}
            \frac{n } {\beta}  \frac{ | \Omega_{1} | }{1 + \Omega_{1} ^ {2}} \frac {|A_{1}H_{1}|} {    \sqrt { 1 - |A_{1}H_{1}|^{2}}  }
\end{eqnarray}
For $|A_{1} H_{1}| \gg 1$ the first term does not go to zero.
Given that this time derivative term much larger than the frequency derivative term, an approximation to the norm is obtained by assuming that the latter term is negligible and hence
\begin{eqnarray}
\nonumber
  \|  \mathbf{\Phi} (\tau , \mu ) \|  &\approx&  \frac { \beta \Omega_{1} |A_{1} H_{1}| \left \{|A_{1} H_{1}| +   \cos (\theta) \right \}}  {1  + |A_{1} H_{1}|^{2}  + 2   |A_{1} H_{1}| \cos (\theta) }.
\end{eqnarray}
This function of angle has minima at $\cos (\theta) = 1$, which are:
\begin{eqnarray}\label{eqn:Gauss_norm_small}
   \|  \mathbf{\Phi} (\tau , \mu ) \| _{\mathrm{min}} &\approx& \frac{  \beta |\Omega_{1}| |A_{1} H_{1}|  }{|A_{1} H_{1}| +  1 }
\end{eqnarray}
Together (\ref{eqn:approx_gamma}) and (\ref{eqn:Gauss_norm_small}) approximate the behaviour of the minima of the norm $\|  \mathbf{\Phi} (\tau , \mu ) \| $ as a function of the frequency separation $\Omega_{1}$ and the amplitude of the the second tone as observed at the output to the filter $|A_{1}H_{1}|$.  When two tones are present the norm will not go to zero. However, despite that,  the test (\ref{eqn:stationary_phase_test}) may be satisfied for a given threshold $C_{1}$ twice per period of the beat frequency at $ \cos (\theta) = - |A_{1}H_{1}|$ and  thus indicate that a region of stationary phase is present.
It is also evident from (\ref{eqn:approx_gamma}) and (\ref{eqn:Gauss_norm_small}) that the behaviour of the norm changes significantly at the point where $|A_{1}H_{1}|=1$.

\begin{figure}
    \centering
        \includegraphics[width=0.9\linewidth]{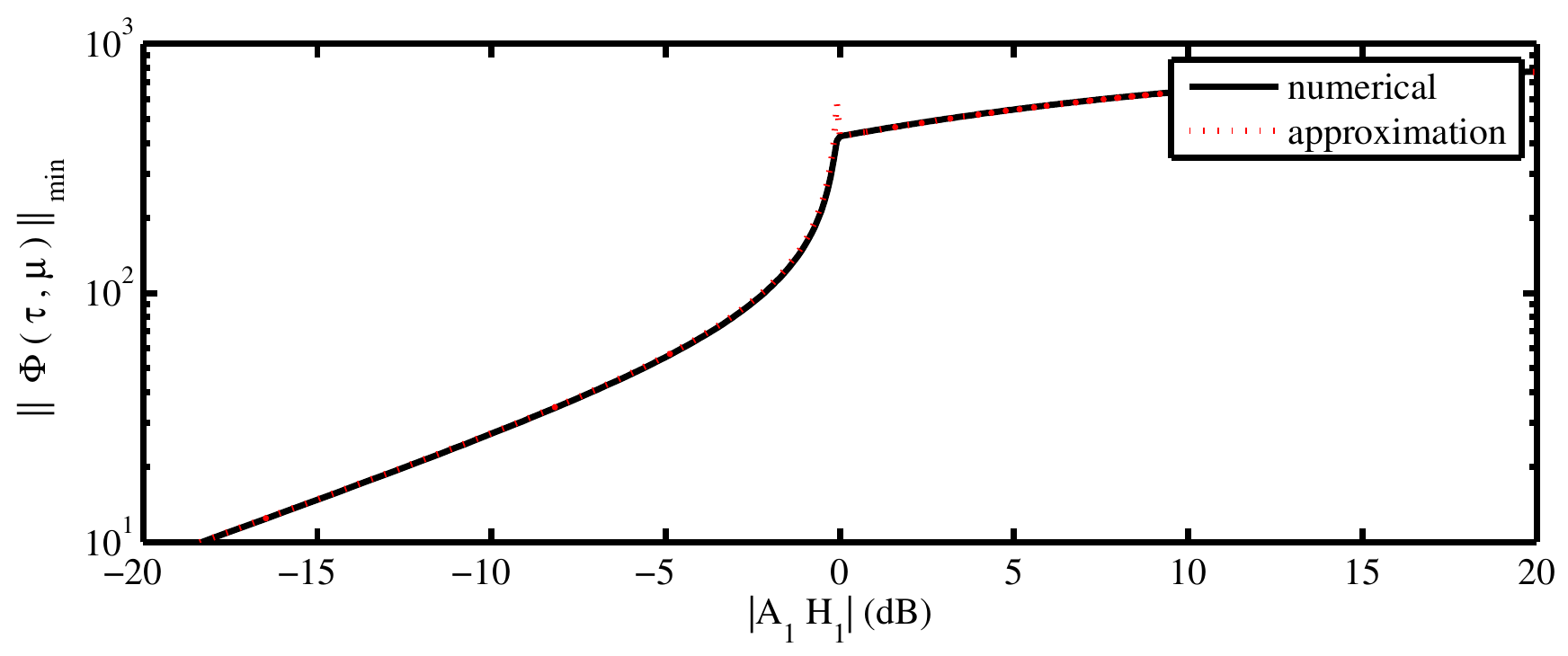}
    \caption{Numerical and approximate evaluation of minimum of phase rate norm of cochlear spaced gammatone filter: filter frequency 1 kHz and $\Omega_{1} = 1$ }
    \label{fig:knee}
\end{figure}

\begin{figure}
    \centering
        \includegraphics[width=0.9\linewidth]{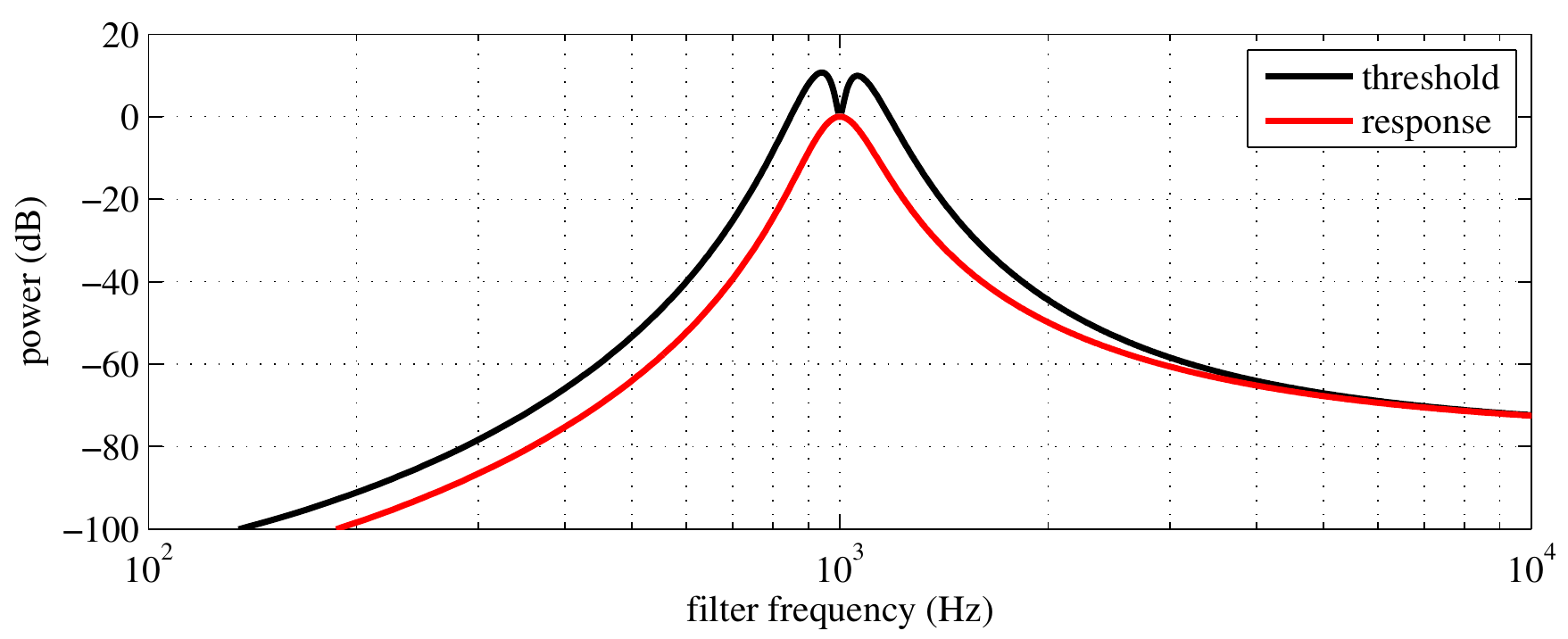}
    \caption{Detection threshold of gammatone cochlear-spaced filter bank for $C_{1} = 10$ ; 0 dB masking tone at 1 kHz; response of filterbank to masker also shown for reference}
    \label{fig:threshold_mask}
\end{figure}

 Figure \ref{fig:knee} illustrates $\|  \mathbf{\Phi} (\tau , \mu ) \| _{\mathrm{min}} $ as a function of $|A_{1}H_{1}|$ for a particular filter and a frequency separation, $\Omega_{1} = 1$. The minimum value of the norm is evaluated both by numerical minimization of (\ref{eqn:gamma_norm_min_monty2}) and using the approximations of (\ref{eqn:approx_gamma}) and (\ref{eqn:Gauss_norm_small}). There is a clear discontinuity at $|A_{1}H_{1}| = 1$ as might be expected from the development of the approximations. The approximation are generally a good fit to the minima evaluated numerically. For this example the most significant errors can be observed around $|A_{1}H_{1}| = 1$.
It is also clear from the numerical evaluation that the minimum of the norm is a montotonically increasing function of $|A_{1}H_{1}|$ and as such is invertible. Thus if a particular value of the threshold $C_{1}$ in (\ref{eqn:stationary_phase_test}) is chosen to detect stationary phase points a corresponding value of $A_{1}H_{1}$ can  be calculated that will just achieve a minimum of $C_{1}$. Together (\ref{eqn:approx_gamma}) and (\ref{eqn:Gauss_norm_small}) are not guaranteed to provide a montonically increasing function. However minor adjustments in the vicinity of the discontinuity can overcome this and together they provide a simple means of inverting the function. Figure \ref{fig:threshold_mask} illustrates how adoption of the stationary phase test (\ref{eqn:stationary_phase_test}) leads to a simultaneous masking effect. A 0 dB masking tone at 1 kHz is introduced that corresponds to the interferer at $\omega_{1}$ rad/s. For each filter frequency $\omega$ rad/s,  (\ref{eqn:approx_gamma}) and (\ref{eqn:Gauss_norm_small}) are used to calculate the amplitude of a tonal component at $\omega$ rad/s that would produce a value of minimum norm equal to the threshold value of $C_{1}$. This amplitude is converted to deciBels to give the detection threshold. For reference, the response of each of the filters, i.e. $|X_{\omega} (\tau)|$, to the masking tone alone is shown. When $\omega_{1} = \omega$ a single tone is present and the results of sub-section \ref{sec:elementary_phasor} apply. The norm will go to zero at $\omega_{1}$ and thus the test will be satisfied. There is a region around $\omega_{1}$ where the detection threshold increases as $|\omega - \omega_{1}|$ increases. However outwith that region the detection threshold follows the same trends as $|X_{\omega} (\tau)|$ and, as such, this masking effect is not symmetrical and affects filters above $\omega_{1}$ more than below it.

\section{Conclusions}
\label{ref:conclusions}
The starting point for this paper was an examination of the application of the PSP to non-asymptotic integrals in general and TF synthesis in particular. The conclusion was that only one aspect of the PSP, location of stationary phase points, is required. The second aspect, approximation of the integral through use of the second derivative of the phase of the integrand, is only needed when closed form expressions that approximate the integral are required.
When this requirement is removed, the second aspect can be
replaced with a test for phase rate dominance. Regions of the TF plane that pass the test and do not contain stationary phase points contribute little or nothing to the final output. Analysis values that lie in these regions can thus be set to zero. In regions of the TF plane that fail the test or in the vicinity of stationary phase points, synthesis is performed in the usual way. In re-examining the application of the PSP to the TF synthesis integral, a new interpretation of the location parameters associated with the synthesis filters leads to: (i) a test for locating stationary phase points in the TF plane; (ii) a test for phase rate dominance in that plane.
With this formulation the stationary phase regions of several elementary signals have been identified theoretically and
it has been shown that sparse reconstructions of tones and chirps are possible.
An analysis of the TF phase rate characteristics  for the case of two simultaneous tones predicts and quantifies a form of simultaneous masking similar to that which characterizes the auditory system. All of the above was addressed from the perspective of deterministic signals. Consideration of random processes is left for future work.

\appendices
%
\section{}
\label{app:proto}
This appendix details the relevant characteristics of Gaussian,
gammatone and gammachirp prototype filters. All have a maximum unit gain at
zero frequency. These prototype filters are scaled in time and
modulated in frequency to form uniformly-spaced, logarithmically-spaced and cochlear-spaced analysis and synthesis filter banks.
A \textit{Gaussian pulse} has a non-causal impulse response $  h(t) = \frac{1}{\sqrt{2\pi}} \mathrm{e}^{-t^{2} /2}$
with  a peak at $t=0$, a group delay of $0$ and a derivative $ \frac{dh(t)}{dt}= -t h(t)$.
It's frequency response is $H(\Omega)= \mathrm{e} ^{- \frac{\Omega ^{2}}{2}}$ with derivative $ \frac{dH(\Omega)}{d \Omega}=  -\Omega H( \Omega )$.
An order $n$ \textit{gammachirp pulse}, with chirp rate parameter $c$ has a causal impulse response defined in terms of the complex gamma function $\Gamma(.)$ as
\begin{eqnarray}
\nonumber
  h(t) & \triangleq&  \frac{ \{ 1 + \jmath \frac{c}{n} \} ^{n+\jmath c}}{\Gamma (n + \jmath
  c)} t ^{n-1} \mathrm{e} ^{  - \{ 1 + \jmath \frac{c}{n} \} t + \jmath c \ln (t) } , \,\,\, \forall t \geq 0,
\end{eqnarray}
$h(t) = 0$, otherwise.  This is a minor modification to the gammachirp filter of \cite{Irino1997JASA} in order to decouple the dependency between the location of the peak gain in the frequency response and the chirp rate parameter $c$. The time derivative is:
\begin{eqnarray}\label{eqn:gamma_chirp_grad}
\nonumber
  \frac{d h (t)}{dt}
    &=&  \left \{ -1 + \frac{n-1}{t} + \jmath  c \left \{  \frac{1}{t} -   \frac{1}{n} \right \} \right \} h(t)
\end{eqnarray}
The gammatone pulse is a specific case of this when the chirp rate parameter is zero, i.e. $c=0$. The gammachirp pulse has frequency response:
\begin{eqnarray}
\nonumber
   H(\Omega)&=& \frac{\left \{ \jmath \frac{c}{n}  +1 \right \}^{n+\jmath c}}{\left \{ \jmath \left \{ \Omega + \frac{c}{n} \right \} +1 \right \}^{n+\jmath c}}
\end{eqnarray}
The maximum gain of the frequency response is 0 dB which occurs at $ \Omega = 0$.
The derivative of its frequency response is
\begin{eqnarray}
\nonumber
   \frac{d H(\Omega)}{d \Omega}
\nonumber
    &=& \frac{ \{ c - \jmath n \}  }{\left \{ \jmath \left \{ \Omega + \frac{c}{n} \right \} +1 \right \}} H(\Omega)
   \end{eqnarray}
The peak in the magnitude of the impulse response occurs at
$t=n-1$ while the group delay at $\Omega = 0$ is $
-\frac{\partial\angle H(0)}{\partial\Omega} = n $.
The gammachirp pulse, like the Gaussian, is scaled (in time) and modulated to give the analysis filter of \ref{eqn:filter}.
The chirp rate parameter is unaffected by the time scaling because of the action of the logarithm function i.e. $\jmath c \ln (\beta t) = \jmath c \ln(\beta) + \jmath c \ln(t)$. The time scaling adds a phase shift $c \ln(\beta)$ to the impulse response. Thus values of $c$ can be used interchangeably with \cite{Irino1997JASA}.
From (\ref{eqn:gamma_chirp_grad}), the phase derivative of the impulse response is:
$  \frac{d \angle h(t)}{d t} =   -\frac{c}{n} + \frac{c} {t}$,
which is zero at the group delay of $t=n$. Likewise $  \frac{d  h(t)}{d t} =   -\frac{1}{n} $ at the group delay.

\section{}
\label{app:PSFNAI2D}
\setcounter{equation}{52}
Consider a complex function $f(\mathbf{v}) = a ( \mathbf{v} ) \mathrm{e}^{\jmath b(\mathbf{v})}$ of a vector
$\mathbf{v} = [v_{1} \, \, v_{2} ]^\mathrm{T}$ with $a, b, v_{1}, v_{2} \in \mathbb{R}$.
The gradient vector is $\boldsymbol{\nabla} _{f} \triangleq [ \frac{\partial f}{\partial v_{1}} \,\, \frac{\partial f}{\partial v_{2}}  ]^\mathrm{T}$. Thus
\begin{eqnarray}
\nonumber
   \frac{\boldsymbol{\nabla} _{f} (\mathbf{v}) }{f (\mathbf{v})} &=&  \frac{1}{a(\mathbf{v})} \left [                                                                \begin{array}{c}
   \frac{\partial a(\mathbf{v})}{\partial v_{1}} \\
   \frac{\partial a(\mathbf{v})}{\partial v_{2}} \\
   \end{array}
   \right ] + \jmath  \left [                                                                \begin{array}{c}
   \frac{\partial b(\mathbf{v})}{\partial v_{1}} \\
   \frac{\partial b(\mathbf{v})}{\partial v_{2}} \\
   \end{array}
   \right ]\\
   &\triangleq& \boldsymbol{\nabla}_{a}(\mathbf{v}) + \jmath  \boldsymbol{\nabla}_{b}  (\mathbf{v})
\end{eqnarray}
The integral of the function around a point $\mathbf{v}_{0}$ over an interval $S = \{ \mathbf{v}: \| \mathbf{v} \| < r \}$  with radius $r$ is  $I_{S} (\mathbf{v}_{0}) = \int _{S} f (\mathbf{v}_{0} + \mathbf{v}) \mathrm{d} \mathbf{v}$. Assuming that the function is well approximated by its first order Taylor series over this interval and making a change of variable $\mathbf{v} = \mathbf{Q} \mathbf{u}$, where $\mathbf{Q}$ is a orthonormal rotation matrix chosen such that
$  \mathbf{Q}^{T} \boldsymbol{\nabla}_{a}(\mathbf{v}_{0}) =
  \left[\,
   \| \boldsymbol {\nabla}_{a} (\mathbf{v}_{0})\|                                                                       \, \, 0
   \, \right] ^{T} $
and
$  \mathbf{Q}^{T}\boldsymbol{\nabla}_{b}  (\mathbf{v}_{0}) \triangleq \left[ \,                                                                   \nabla_{b1} (\mathbf{v}_{0}) \,\, \nabla_{b2} (\mathbf{v}_{0})                                                           \, \right] ^{T}$,
gives:
\begin{eqnarray}
\nonumber
\lefteqn{  I_{S} (\mathbf{v}_{0}) } \\
\nonumber
   &\approx&  f( \mathbf{v}_{0})  \int _{S} \left \{ 1 + \mathbf{u}^{T} \mathbf{Q}^{T} \boldsymbol{\nabla}_{a}(\mathbf{v}_{0}) + \jmath \mathbf{u}^{T} \mathbf{Q}^{T}\boldsymbol{\nabla}_{b}  (\mathbf{v}_{0}) \right \}  \mathrm{d} \mathbf{u} \\
\nonumber
   &=& f( \mathbf{v}_{0}) \left \{  \iint\limits_{S} \mathrm{d} u_{1} \mathrm{d} u_{2} \right.  \\
\nonumber
    & & \,\, + \,  \left \{ \| \boldsymbol {\nabla}_{a} (\mathbf{v}_{0})\| + \jmath \nabla_{b1} (\mathbf{v}_{0}) \right \}  \iint\limits_{S} u_{1} \mathrm{d} u_{1} \mathrm{d} u_{2} \\
\nonumber
    & & \,\, + \, \left. \jmath \nabla_{b2} (\mathbf{v}_{0})  \iint\limits_{S} u_{1} \mathrm{d} u_{1} \mathrm{d} u_{2} \right \}
\end{eqnarray}
Given that the interval is circular $\iint\limits_{S} u_{1} \mathrm{d} u_{1} \mathrm{d} u_{2}  = \iint\limits_{S} u_{1} \mathrm{d} u_{2} \mathrm{d} u_{2} $, the integral becomes
\begin{eqnarray}
\lefteqn{ I_{S} (\mathbf{v}_{0})   \approx  f( \mathbf{v}_{0}) \left \{  \iint\limits_{S} \mathrm{d} u_{1} \mathrm{d} u_{2} \right. } \hspace{3.5in} \nonumber \\
\nonumber
\lefteqn{      \left. \,\, + \,  \left \{ \| \boldsymbol {\nabla}_{a} (\mathbf{v}_{0})\| + \jmath \left \{ \nabla_{b1} (\mathbf{v}_{0}) + \nabla_{b2} (\mathbf{v}_{0}) \right \} \right\} \iint\limits_{S} u_{1} \mathrm{d} u_{1} \mathrm{d} u_{2} \right \} }\hspace{3.5in}
\end{eqnarray}
By definition, the rotation matrix is given by, $\mathbf{Q}^{T}  = \left[\begin{smallmatrix}
  \cos (\theta) & \sin(\theta) \\
  -\sin(\theta) & \cos(\theta)
 \end{smallmatrix}
 \right]
$,
where$    \left[
     \begin{array}{c}
       \cos (\theta) \\
       \sin(\theta) \\
     \end{array}
   \right]
   = \frac{\boldsymbol{\nabla}_{a}(\mathbf{v}_{0})}{\|\boldsymbol{\nabla}_{a}(\mathbf{v}_{0}) \|}.$
Thus the integral over the interval will be dominated by the phase variations if $| \nabla_{b1} (\mathbf{v}_{0}) + \nabla_{b2} (\mathbf{v}_{0}) | \gg\| \boldsymbol {\nabla}_{a} (\mathbf{v}_{0})\|$, where
\begin{eqnarray}
\nonumber
   \nabla_{b1} (\mathbf{v}_{0}) + \nabla_{b2} (\mathbf{v}_{0}) &=& \left[
                                                                     \begin{array}{cc}
                                                                       1 & 1 \\
                                                                     \end{array}
                                                                   \right]
  \mathbf{Q}^{T}\boldsymbol{\nabla}_{b}  (\mathbf{v}_{0}) \\
\nonumber
&=&    \frac{\boldsymbol{\nabla}_{a}^{T}(\mathbf{v}_{0})}{\|\boldsymbol{\nabla}_{a}(\mathbf{v}_{0}) \|}
\mathbf{W}
\boldsymbol{\nabla}_{b}  (\mathbf{v}_{0})
\end{eqnarray}
and $\mathbf{W}   = \left[
  \begin{smallmatrix}
      1 & 1 \\
    -1 & 1
  \end{smallmatrix}
\right]$.
Therefore a test for first order phase rate dominance is
\begin{eqnarray} \label{eqn:phase_rate_dominance_app}
   \frac{|\boldsymbol{\nabla}_{a}^{T}  (\mathbf{v}_{0}) \mathbf{W} \boldsymbol{\nabla}_{b}(\mathbf{v}_{0})  | }{\| \boldsymbol{\nabla}_{a}(\mathbf{v}_{0})\|} &\gg& \| \boldsymbol{\nabla}_{a}(\mathbf{v}_{0})\|.
\end{eqnarray}


\bibliography{paper_aug2012}


\end{document}